\documentclass[5p, twocolumn]{elsarticle}
\journal{Journal of Systems and Software}

\usepackage{amsmath,amssymb,amsfonts} % Math symbols
\usepackage{graphicx}                 % Images
\usepackage{array, multirow}          % Advanced table formatting
\usepackage{enumitem}                % Custom lists
\usepackage{float}                   % Float control
\usepackage{dblfloatfix}
\usepackage{flafter}
\usepackage{flushend}
\usepackage{caption}                 % Caption customization
\usepackage{listings}                % Code listings
\usepackage[hyphens]{url}                     % URL support

\usepackage{verbatim}                % Comment blocks
\usepackage{hyphenat}                % Hyphenation control
\usepackage[colorlinks=true,breaklinks=true]{hyperref} % enable clickable and breakable links

\hypersetup{
    breaklinks=true,     % allow links to break over lines
    colorlinks=false,
    linkcolor=blue,
     citecolor=green,
     urlcolor=blue
}

\usepackage[dvipsnames]{xcolor}      % Extended color support
\usepackage[export]{adjustbox}       % Image alignment
\usepackage[many]{tcolorbox}         % Color boxes

\usepackage{fontawesome}             % Icons

\usepackage{xparse}                  % Advanced macros
\usepackage{etoolbox}               % Logic tools for macros

\usepackage{hyperref}               % Hyperlinks

\usepackage{todonotes}              % \todo command

\usepackage{soul}                   % Underlining
\setuldepth{Berlin}                % Underline depth tuning

\usepackage[resetlabels,labeled]{multibib} % Multiple bibliographies

\newcommand{\wlcite}[1]{%
  \def\lastcomma{}%
  \forcsvlist{\osformat}{#1}%
}

\newcommand{\glcite}[1]{%
  \def\lastcomma{}%
  \forcsvlist{\osformath}{#1}%
}
\newcommand{\whiteLitteraturePrefix}{WL}
\definecolor{whiteLitteratureColor}{HTML}{4870B7}    % Setting main color
\newcommand{\grayLitteraturePrefix}{GL}
\definecolor{grayLitteratureColor}{HTML}{807F80}     % Setting sub color

\newcites{WL}{Selected White Literature}
\newcites{GL}{Selected Gray Literature}

\makeatletter

\newcommand{\mycitenumWL}[2][]{%
  \begingroup
    \def\@citex##1##2##3{##1}%
    \ifstrequal{#1}{highlight}{\textcolor{grayLitteratureColor}{\citeWL{#2}}}{\textcolor{whiteLitteratureColor}{\citeWL{#2}}}%
  \endgroup
}

\newcommand{\mycitenumGL}[2][]{%
  \begingroup
    \def\@citex##1##2##3{##1}%
      \ifstrequal{#1}{highlight}{\textcolor{grayLitteratureColor}{\citeGL{#2}}}{\textcolor{whiteLitteratureColor}{\citeGL{#2}}}%
 \endgroup
}

\makeatother
\let\oldciteWL\citeWL
\renewcommand{\citeWL}[1]{\textcolor{whiteLitteratureColor}{\oldciteWL{#1}}}

\let\oldciteGL\citeGL
\renewcommand{\citeGL}[1]{\textcolor{grayLitteratureColor}{\oldciteGL{#1}}}

\newcommand{\osformat}[1]{%
  \ifx\lastcomma\empty
    \textcolor{whiteLitteratureColor}{\whiteLitteraturePrefix}\mycitenumWL{#1}%
    \gdef\lastcomma{notempty}%
  \else
, \textcolor{whiteLitteratureColor}{\whiteLitteraturePrefix}\mycitenumWL{#1}%
  \fi
}

\newcommand{\osformath}[1]{%
  \ifx\lastcomma\empty
    \textcolor{grayLitteratureColor}{\grayLitteraturePrefix}\mycitenumGL{#1}% manually add [highlight] if needed
    \gdef\lastcomma{notempty}%
  \else
, \textcolor{grayLitteratureColor}{\grayLitteraturePrefix}\mycitenumGL{#1}%
  \fi
}

\definecolor{main}{HTML}{CFCFCF}    % Setting main color
\definecolor{sub}{HTML}{CFCFCF}     % Setting sub color

\tcbset{
    sharp corners,
    colback = white,
    before skip = 0.5cm,
    after skip = 0.7cm
}

\newtcolorbox{boxCNoTitle}{
    top=10pt,
    rounded corners,
    coltitle=black,
    colframe=gray,
    colbacktitle=sub,
    colback = sub,
    enhanced,
    center,
    boxrule=0.5pt
}

\newtcolorbox{boxC}[2][]{aibox,title=#2,#1}

\tcbset{
  aibox/.style={
    top=10pt,
    boxrule=0.5pt,
    rounded corners,
    coltitle=black,
    colframe=gray,
    colbacktitle=sub,
    colback = sub,
    enhanced,
    center,
    attach boxed title to top left={yshift=-0.1in,xshift=0.15in},
    boxed title style={boxrule=0.5pt,colframe=gray,},
  }
}

\newcounter{keyTakeAwaysCounter}
\newenvironment{keyTakeAways}[1][Key Take Away]
{
    \addtocounter{keyTakeAwaysCounter}{1}
    \begin{boxC}{\faLightbulbO ~ \thekeyTakeAwaysCounter. #1}
}{
    \end{boxC}
}

\newcounter{keyRQAnswerCounter}
\newcounter{keyLimitationsCounter}
\newcommand{\ReviewerA}[1]{\textcolor{black}{#1}}
\newcommand{\ReviewerB}[1]{\textcolor{black}{#1}}
\newcommand{\ReviewerC}[1]{\textcolor{black}{#1}}

\newenvironment{ReviewerAEnv}{\color{black}}{}

\newenvironment{ReviewerCEnv}{\color{black}}{}

\usepackage{tikz}
\usetikzlibrary{positioning,arrows.meta}

\colorlet{colValidation}{blue}
\colorlet{colEthicalAI}{red}
\colorlet{colOversight}{green!60!black}
\colorlet{colSecurity}{orange!80!black}
\colorlet{colStudies}{purple}
\colorlet{colSALC}{teal}
\colorlet{colFrameworks}{brown}
\colorlet{colMulti}{magenta!80!black}

\begin{document}
\begin{frontmatter}

\title{Generative AI for Software Architecture. \\ Applications, Challenges, and Future Directions}

\author[OULU]{Matteo Esposito}
\ead{matteo.esposito@oulu.fi}
\author[OULU]{Xiaozhou Li}
\ead{xiaozhou.li@oulu.fi}
\author[OULU,TUNI]{Sergio Moreschini}
\ead{sergio.moreschini@oulu.fi}
\author[OULU]{Noman Ahmad}
\ead{noman.ahmad@oulu.fi}
\author[ARIZ]{Tomas Cerny}
\ead{tcerny@arizona.edu}
\author[IIIT]{Karthik Vaidhyanathan}
\ead{karthik.vaidhyanathan@iiit.ac}
\author[OULU]{Valentina Lenarduzzi}
\ead{valentina.lenarduzzi@oulu.fi}
\author[OULU]{Davide Taibi}
\ead{davide.taibi@oulu.fi}

\address[OULU]{University of Oulu, Finland}
\address[TUNI]{Tampere University, Finland}
\address[ARIZ]{University of Arizona, USA}
\address[IIIT]{Software Engineering Research Center, IIIT Hyderabad, India}

\begin{abstract}
\noindent\textbf{Context}. Generative Artificial Intelligence (GenAI) is transforming much of software development, yet its application in software architecture is still in its infancy. 

\noindent\textbf{Aim}. Systematically synthesize the use, rationale, contexts, usability, and future challenges of GenAI in software architecture.

\noindent\textbf{Method}. Multivocal literature review (MLR), analyzing peer-reviewed and gray literature, identifying current practices, models, adoption contexts, reported challenges, and extracting themes via open coding.

\noindent\textbf{Results}: This review identifies a significant adoption of GenAI for architectural decision support and architectural reconstruction. OpenAI GPT models are predominantly applied, and there is consistent use of techniques such as few-shot prompting and retrieved-augmented generation (RAG). GenAI has been applied mostly to the initial stages of the Software Architecture Life Cycle (SALC), such as Requirements-to-Architecture and Architecture-to-Code. Monolithic and microservice architectures were the main targets. However, rigorous testing of GenAI outputs was typically missing from the studies. Among the most frequent challenges are model precision, hallucinations, ethical aspects, privacy issues, lack of architecture-specific datasets, and the absence of sound evaluation frameworks.

\noindent\textbf{Conclusions}: GenAI shows significant potential in software design, but there are several challenges on its way toward greater adoption. Research efforts should target designing general evaluation methodologies, handling ethics and precision,  increasing transparency and explainability, and promoting architecture-specific datasets and benchmarks to overcome the gap between theoretical possibility and practical use.
\end{abstract}


% Matteo: Do not touch :D
% \let\oldciteOS\citeOS
% \renewcommand{\citeOS}[1]{OS\oldciteOS{#1}}

% \let\oldciteWL\citeWL
% \renewcommand{\citeWL}[1]{\textcolor{whiteLitteratureColor}{WL\oldciteWL{#1}}}

% \let\oldciteGL\citeGL
% \renewcommand{\citeGL}[1]{\textcolor{grayLitteratureColor}{WL\oldciteGL{#1}}}

\end{frontmatter}

\section{Introduction}
\label{sec:intro}
Generative AI (GenAI) is driven by the need to create, innovate, and automate complex tasks that traditionally require human creativity. It empowers companies and individuals to unlock new possibilities, promote innovation, and improve productivity~\cite{esposito_beyond_2024}.

In software engineering, GenAI is revolutionizing the way developers design, write, and maintain code~\cite{russo2024navigating}.  Given its potential and benefits, the integration of GenAI within the domain of software engineering has gained increasing attention as it has a transformative potential to enhance and automate various aspects of the software development lifecycle \cite{sauvola2024future}. 

Although GenAI has shown its capabilities in areas such as code generation, software documentation, and software testing \cite{Jahic24,beyond_2024}, its application in software architecture remains an emerging area of research, with ongoing debates about its effectiveness~\cite{Dhar24a}, reliability~\cite{Raghavan24}, and best practices~\cite{Soliman25}. Researching the application of GenAI in software architecture is crucial because it has the potential to transform the way complex systems are designed, optimized, and maintained. 

However, practitioners and researchers continue to be challenged in understanding the implications, limitations, and potential benefits of GenAI for architectural tasks. To catalyze research in this area, they need a roadmap on various research directions, applications, trends, challenges, and future directions.

% \todo[inline]{Tomas: as a reader, I am not excited here. So I did a lot of tweaks above, is it motivating now?}
% \todo[inline]{Davide: Way better, thank you :) }

To better understand the existing research in this area, we \textbf{investigated the current state of research and practice on the use of GenAI in software architecture}. 

Specifically, we conducted a Multivocal Literature Review (MLR) to synthesize the findings from academic literature and gray literature sources, including industry reports, blog posts, and technical documentation~\cite{MLRguidelines}. 
In particular, our goal is to understand how GenAI is used in software architecture and what the underlying rationales, models, and usage approaches are, as well as the context and practical use cases where GenAI has been adopted for software architecture. Moreover, we aim to understand research gaps highlighted by the literature and to provide an overview of possible research directions to practitioners and researchers. 

Despite the growing adoption of GenAI in software engineering, several factors justify the need for a systematic investigation into its role in software architecture:
\begin{itemize}
    \item \textit{Emerging and Underexplored Research Area}: Although GenAI has been widely adopted in software architecture tasks, its role in software architecture remains underdeveloped~\cite{Jahic24}. Studies suggest that while GenAI models can help in architectural modeling and decision-making, their contributions are still in the early stages of research and adoption~\cite{Dhar24a}.
    
    \item \textit{Lack of Systematic Evidence on Effectiveness and Reliability}: Existing work reports inconsistent findings regarding the reliability of GenAI for architectural decisions~\cite{Raghavan24}. Some studies indicate its potential in architectural modeling and automation, while others highlight challenges such as hallucinations, interpretability, and alignment with established architectural principles~\cite{Soliman25}.
    
    \item \textit{Need for a Comprehensive Synthesis of Both Academic and Gray Literature}: Given the rapid evolution of GenAI models, gray literature, such as industry reports and practitioner blogs, provides valuable but fragmented knowledge that needs systematic integration~\cite{kaplan2024combining}.
    
    \item \textit{Unclear Best Practices and Guidelines for Adoption}: Although strategies such as prompt engineering, Retrieval-Augmented Generation (RAG), and fine-tuning have been explored, there is no consensus on best practices for effectively using GenAI in different software architecture tasks~\cite{Diaz24, Diaz25}. A structured review can help identify and formalize these practices for both researchers and practitioners~\cite{Supekar24}.
    
    \item \textit{Increasing Industry Interest in Architectural Automation}: Enterprises are increasingly exploring AI-assisted architectural decision-making tools, yet there is still limited understanding of their practical benefits and risks~\cite{Fujitsu25}. The demand for explainable AI in architecture, and in particular in safety-critical domains, highlights the need for a systematic evaluation of the literature~\cite{Martelli23}.
    
    \item \textit{Identifying Open Challenges}: Multiple research questions remain open on multiple aspects. Examples are security vulnerabilities introduced by AI-driven modifications~\cite{Raghavan24}, biases in architectural decision making~\cite{Johansson24}, or ethical implications of AI-generated architectural decisions~\cite{Eisenreich24}. This work will help illuminate open challenges highlighted by practitioners and researchers. 
\end{itemize}

\ReviewerA{Our results reveal that, while GenAI excels at automating tasks grounded in natural language and structured templates, its integration into complex, high-stakes architectural decision-making remains limited. The prevailing utilization of these tools is predominantly oriented towards documentation and code generation, with a paucity of examples addressing system-level reasoning, trade-off analysis, or performance modeling.
Furthermore, most studies evaluated GenAI tools based on usability or accuracy metrics; few addressed the impact of these tools on architectural quality attributes such as modifiability, scalability, or maintainability. These findings demonstrate an absence of rigorous evaluation.}

The main contributions of this study are as follows.
\begin{itemize}
    \item \textbf{A comprehensive synthesis} of the existing literature and industry reports to provide an overview of how GenAI is used in software architecture.
	\item \textbf{A classification of the GenAI models} adopted for Software Architecture based on data extracted following the open coding approach~\cite{CorbinStrauss2008}. 
    \item \textbf{Identification of Common Applications, benefits, and challenges} of the application of GenAI in software architecture.
	\item \textbf{Identification of research gaps} and open research questions that provide recommendations for future studies and practical adoption.
	\item \textbf{Industry Relevance} By incorporating the gray literature, we bridge the gap between research and practice, ensuring that our findings are aligned with real-world applications.
    %\todo[inline]{Tomas: Some items have ':' some not, should we unify? }

\end{itemize}	
% \todo[inline]{Davide: We  must double check the contribution in the end and synch with the actual results}

\textbf{Paper Structure:}  
Section~\ref{sec:relatedwork} presents the related work. Section~\ref{sec:Methodology} describes the study design. Section~\ref{sec:results} presents the results obtained, and Section~\ref{sec:Discussion} discusses them. Section~\ref{sec:threats} highlights the threats to the validity of our study. Finally, Section~\ref{sec:conclusions} draws the conclusion.

\section{Related Work}
\label{sec:relatedwork}

Different works have been done to understand the extent to which large language models have been applied in software engineering. \citet{fan2023large} performed a survey to identify how LLMs have been leveraged by different steps in the software engineering lifecycle. The work highlights that while much emphasis has been given to implementation, particularly code generation, not much work has been done in the area of using LLMs for requirements and design. This is further emphasized by \citet{hou2024large}, where the authors performed a systematic literature review to understand the usage of LLMs in software engineering with a particular focus on how LLMs have been leveraged to optimize processes and outcomes. The authors analyzed 395 research articles and concluded that similar to the previous study, most of the applications of LLMs have been on software development. It is also important to note that the work only selected four relevant academic literature that leverage LLMs for software design. Thereby emphasizing the need for a multi-vocal literature review. \citet{ozkaya2023application} provided a pragmatic view into using LLMs for Software Engineering tasks by enlisting the opportunities, associated risks, and potential challenges. The work points out challenges such as bias, data quality, privacy, explainability, etc, while describing some of the opportunities with respect to specification generation, code generation, documentation, etc. 

\begin{table*}[t]
\centering
\caption{Classification and Comparison of Related Systematic Studies\\{\scriptsize \textbf{Legend}: \textbf{SLR} - Systematic Literature Review; \textbf{SMS} - Systematic Mapping Study; \textbf{MLR} - Multivocal Litterature Review; \textbf{Hol} Holistic Review}}
\label{tab:related_work_comparison}
\centering
\footnotesize
\begin{tabular}{m{1.5cm} m{1cm} m{3.3cm} m{3cm} m{7cm}}
\hline
\textbf{Reference} & \textbf{Systematic Study Type} & \textbf{Main Focus Area} & \textbf{Identified Challenges} & \textbf{Key Findings} \\
\hline

\citet{hou2024large} & SLR & Process optimization using LLMs & Limited software design applications & Majority use in software development phases, underscoring the need for multi-vocal studies. \\
\hline
\citet{ozkaya2023application} & Hol & Risks and opportunities of LLMs in SE & Bias, data quality, privacy, explainability & Highlights potential in specification, code, and documentation generation tasks. \\
\hline
\citet{jiang2024survey} & SLR & LLMs for code generation & Bridging research-practice gap & Taxonomy developed; outlined research-practice gaps and future opportunities. \\
\hline
\citet{10440574} & SLR & LLM applications in software testing & Integration challenges & Extensive LLM usage in testing highlighted; discussed practical integration barriers. \\
\hline
\citet{marques2024using} & Hol & ChatGPT in requirements engineering & Data accuracy and relevance & Provided a detailed overview of current use, challenges, and identified future directions. \\
\hline
\citet{santos2024impacts} & SLR & Generative AI impact on SE lifecycle & Overemphasis on development/testing phases & Confirmed dominance of development/testing; suggested expansion to other SE phases. \\
\hline
\citet{saucedo24migration} & SMS & AI for migration to microservices & Accuracy of unsupervised learning methods & Highlighted clustering as a prevalent AI technique for migrating monolithic to microservices architecture. \\
\hline
\citet{fan2023large} & Hol & LLMs in SE lifecycle & Limited exploration in requirements/design & Emphasis predominantly on code generation; limited attention to early SE phases. \\
\hline
\ReviewerA{\citet{Bucaioni2025}}&\ReviewerA{SLR} & \ReviewerA{AI integration with Software Architecture} & \ReviewerA{Text notes real-time adaptation, trade-off analysis, etc} & \ReviewerA{Vision for AI in architecture including design automation and diagnostics} \\
\hline
\ReviewerA{\citet{Schmid2025}}& \ReviewerA{SLR} &\ReviewerA{AI usage in Software Architecture} & \ReviewerA{Gaps in code generation from architecture, cloud-native, etc.}	& \ReviewerA{Key application areas; gaps in advanced prompting and evaluation}\\ 
\hline
\textbf{Our Work} & MLR   & Generative AI specifically for software architecture & Scarcity of comprehensive reviews; dominance of grey literature & Provides comprehensive insights, bridging academic and industry perspectives in generative AI applied to software architecture. \\
\hline
\end{tabular}%

\end{table*}

There have also been various secondary studies focusing on the use of LLMs for specific aspects of Software Engineering. For instance, 
~\citet{jiang2024survey} performed a systematic literature review to understand the use of LLMs for code generation. The authors selected and analyzed around 235 articles and developed a taxonomy of LLMs for code generation. Further, the work points out critical challenges and identifies opportunities to bridge the gap between research and practice of using LLMS for code generation. \citet{10440574}, on the other hand, performed a systematic literature review to identify the different types of work that have used LLMs for software testing. It identified and analyzed 102 relevant studies that have used LLMs for software testing from both the software testing and LLMs perspectives. \citet{marques2024using} performed a comprehensive study to understand the application of LLMs (in particular ChatGPT) in requirements engineering. The work highlights the state of use of ChatGPT in requirements engineering and further lists the challenges and potential future work that needs to be performed in this direction. A secondary study to identify the impact of GenAI on software development activities was performed by~\citet{santos2024impacts}. Like other secondary studies on using LLMs for software engineering, this study also highlighted that most of the work has been centered around development and testing.

While to the best of our knowledge, there is a lack of secondary study on the use of GenAI applied to software architecting practices, there have been some work that leverages LLMs for various software architecting practices. 

\citet{Alsayed2024MicroRec} developed MicroRec, an approach that leverages state-of-the-art deep learning techniques and LLMs to recommend microservices to developers. The approach allows developers to search for microservices in service registries using natural language queries. An approach that leverages GenAI, in particular LLMs, to suggest architectural patterns from requirements was proposed by \citet{Gustrowsky24}.  The proposed solution fine-tunes the Llama 2 LLM on a custom dataset of requirements and architectural patterns. The evaluation demonstrated an accuracy of 70\% on the test set. \citet{kaplan2024combining}, on the other hand, proposed an approach that combines knowledge graphs and LLMs to support effective discovery and access to software architecture research knowledge.

Apart from the works that leverage GenAI, particularly LLMs, there have also been works that applied various AI techniques to software architecting processes/practices. \citet{saucedo24migration} performed a systematic mapping study to understand the use of AI for migrating monolithic systems to microservice-based systems. The study identified unsupervised learning, particularly clustering, as one of the most popular AI techniques used for migration based on observations from 22 primary studies.

\ReviewerA{\citet{Bucaioni2025} performed a systematic literature review and forward-looking vision for integrating AI with Software Architecture. Results shows how AI is already being used in architectural tasks like automated design, trade-off analysis, and continuous documentation updates. They identified some aspects that need to be investigated such as being able to adapt in real-time, being able to follow how changes have been made,  to understand why a decision has been made, and to optimize for more than one thing at a time. They also suggests six areas to research: real-time monitoring and self-adaptation, automated documentation, context-aware reasoning, multi-objective optimisation, integrated multi-level diagnostics, and robust benchmarking for practical evaluation. }

\ReviewerA{\citet{Schmid2025} conducted a systematic literature review examining how LLMs are used within software architecture. They considered 18 papers identifying four primary application areas: reference architectures, classification and detection, extraction and generation, and assistant systems. The achieved results show that most approaches leverage decoder-only models like GPT variants, typically using simple prompting techniques.  While LLM-based methods generally perform better than other methods, there are still several areas that are not well researched. These include generating source code from architectural designs, cloud-native architecture, and checking that things meet standards. In the future, we will be looking into ways to improve input strategies, explore more advanced prompting techniques, and make sure that we keep evaluating the technology as it develops.}

Despite the active exploration of LLMs for a variety of software engineering (SE) tasks, particularly code generation, testing, requirements engineering, etc, there is a dearth of a comprehensive literature review dedicated to LLM for software architecture. Further, many of the works related to using GenAI for software design or software architecture are more available in the grey literature. Hence, in this work, we performed a multi-vocal literature review to identify the existing landscape of using GenAI for software architectural practices and processes.

\section{Methodology}
\label{sec:Methodology}
\begin{figure}
    \centering
    \includegraphics[width=0.8\linewidth]{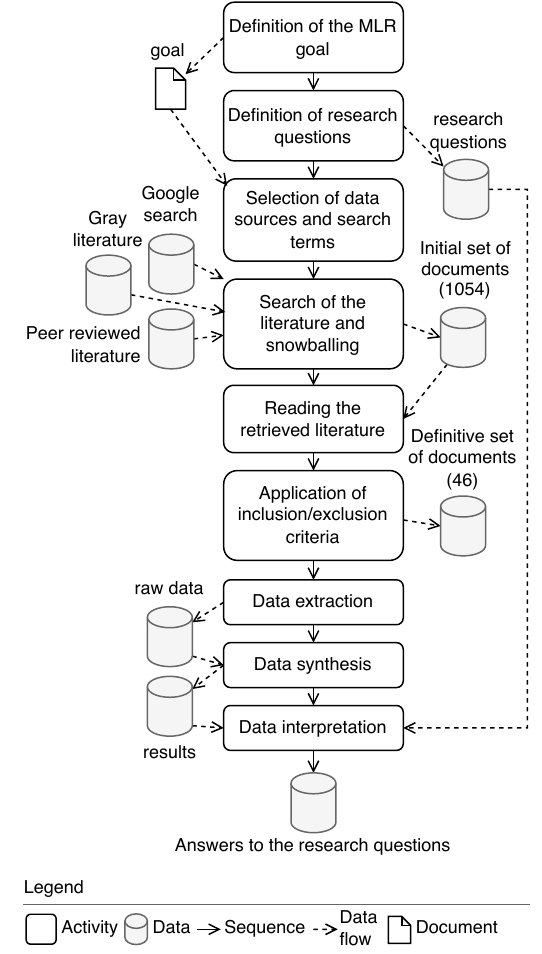}
    \caption{Study Workflow}
    \label{fig:studyworkflow}
\end{figure}

This section addresses the methodology, defining the goal and research questions. It also provides the search and selection process, as well as inclusion and exclusion criteria for both peer-reviewed and gray literature. Our search strategy is presented in Figure~\ref{fig:studyworkflow}.

\subsection{Goal and Research Questions}
The goal of this MLR is to provide a comprehensive overview of GenAI's role in software architecture, from its current state to its prospects. 
We aim to contribute significantly to the body of knowledge in software engineering, providing actionable insights to researchers and practitioners.
To carry out this research, we conducted a multivocal review of the literature \cite{MLRguidelines}.
Based on the objectives of our study, we defined the following research questions (RQs). 

\begin{boxC}{\textbf{RQ$_1$}} 
How is Generative AI utilized in software architecture, and what are the underlying rationales, models, and usage approaches?

\begin{itemize} [leftmargin=*]
    \item \textbf{RQ$_{1.1}$}. (\textit{Why}) For what purposes are Generative AI models used in software architecture?
    \item \textbf{RQ$_{1.2}$}. (\textit{What}) Which Generative AI models have been used?
    \item \textbf{RQ$_{1.3}$}. (\textit{How}) How has Generative AI been applied?
\end{itemize}
\end{boxC}

In this RQ, we aim to investigate the integration of GenAI technologies in the domain of software architecture to highlight the motivations behind the adoption of these technologies, the specific models that have been employed, and the practical applications in software architecture.
We try to understand the underlying rationale behind the adoption of AI models and how they contribute in practice to architectural design, maintenance, and process optimization (\textbf{RQ$_{1.1}$}). 
Therefore, researchers and practitioners can better assess the impact and potential of GenAI in their specific contexts.

However, in-depth investigation of the adopted GenAI models can provide a catalog of the technologies that have been implemented, providing a detailed landscape of the tools available to software architects (\textbf{RQ$_{1.2}$}). 

Other important aspects to be considered are the strategies for implementing GenAI technologies in architectural practices, focusing on the types of projects that benefit from them, and the outcomes of these integrations (\textbf{RQ$_{1.3}$}). 

\begin{boxC}{\textbf{RQ$_2$}}
In what contexts is Generative AI used for software architecture?

\begin{itemize} [leftmargin=*]
    \item \textbf{RQ$_{2.1}$}. (\textit{Where}) In which software \ReviewerA{architecture life cycle phase} is Generative AI applied?
    \item \textbf{RQ$_{2.2}$}. (\textit{For what}) Which architectural \ReviewerB{styles or patterns} are targeted?
    \item \textbf{RQ$_{2.3}$}. (\textit{For what}) Which architectural maintenance tasks and quality-related \ReviewerB{activities} are targeted?
    \item \textbf{RQ$_{2.4}$}. Which architectural analysis or modeling methods have been used to validate Generative AI outputs?
    \item \ReviewerB{\textbf{RQ$_{2.5}$}}. To which use cases has Generative AI been applied in Software Architecture?
\end{itemize}
\end{boxC}

Once GenAI technologies have been investigated in the domain of software architecture, the next step is to explore the environments and scenarios where GenAI is integrated, mapping the conditions or settings in which these technologies are applied. 
Therefore, researchers and practitioners could better identify opportunities where GenAI can be used effectively, improving the architectural design process and addressing complex challenges.
In particular, we identified the stages of the software architecture life cycle where GenAI tools are the most beneficial, such as requirements, design, implementation, testing, or maintenance, providing insight for the continuous integration of AI throughout the development life cycle (\textbf{RQ$_{2.1}$}). 
Another important aspect is to specify for which architectural styles or design patterns (e.g., microservices, monolithic architectures) (\textbf{RQ$_{2.2}$}) a GenAI model is more effective and advantageous in improving design coherence and system scalability (\textbf{RQ$_{2.3}$}). Moreover, since the benefit of adopting a new model should always be validated, it is necessary to evaluate and validate the results produced by GenAI, and architectural analysis or modeling methods have been used  (\textbf{RQ$_{2.4}$}).
Exploring the environments and scenarios where GenAI is integrated \ReviewerA{in architectural task} led to identifying use cases where it has been implemented to highlight versatility and adaptability in different cases to solve specific problems, contribute to innovation, and drive industry advancements (\textbf{RQ$_{2.5}$}).

% \begin{boxC}{\textbf{RQ$_3$}}
% To which use cases has Generative AI been applied \ReviewerA{in Software Architecture}?			
% \end{boxC}

\begin{boxC}{\textbf{RQ$_3$}}
What future challenges are identified for the use of Generative AI in software architecture?		
\end{boxC}

As a last RQ, we investigate the future challenges of GenAI in software architecture for which researchers and practitioners should work in the next years (\textbf{RQ$_3$}). 

\subsection{Search Strategy}
In this Section, we report the process we adopted for collecting the peer-reviewed papers and the gray literature contributions to be included in our revision.

\subsubsection{Search Terms} 
The search string contained the following search terms: 

\begin{boxC}{\textbf{Search String}}
\centering
\textit{(``generative AI'' OR ``gen AI'' OR gen-AI OR genAI OR ``large language model*'' OR ``small language model*'' OR LLM OR LM OR GPT* OR Chatgpt* OR Claude* OR Gemini* OR Llama* OR Bard* OR Copilot OR Deepseek) \\
\textbf{AND} \\
(``software *architect*'' OR ``software design*'' OR ``software decompos*'' OR``software structur*'')}
\end{boxC}

In our search string, we used different terms for GenAI, such as gen AI, gen-AI, or genAI, to increase research efficiency.  We used an asterisk character (*), such as software architect*,  to get all possible term variations, such as plurals and verb conjugations. To increase the likelihood of finding papers that addressed our goal, we applied the search string to the title and abstract.

\subsubsection{Bibliographic Sources} 
\begin{ReviewerAEnv}
    For retrieving the peer-reviewed paper, we selected the list of relevant bibliographic sources following Kitchenham and Charters' recommendations ~\cite{Kitchenham2007} since these sources are recognized as the most representative in the software engineering domain and are used in many reviews. For the white literature, we used four digital libraries: \textit{ACM Digital Library, IEEEXplore Digital Library, Scopus, Web of Science}. Concerning the gray literature, we used 3 search engines: Google, Google Scholar, and Bing ~\cite{MLRguidelines}. 
\end{ReviewerAEnv}

\subsubsection{Inclusion and Exclusion Criteria}
We defined the inclusion and exclusion criteria to be applied to the title and abstract (T/A), the full text (F), or both cases (All), as reported in Table~\ref{tab:Criteria}.

\begin{table}
\centering
\footnotesize
\caption{Inclusion and Exclusion Criteria} 
\label{tab:Criteria} 
\begin{tabular}
{@{}p{0.2cm}|p{6.5cm}|p{0.5cm}@{}}
\hline
\textbf{ID} & \textbf{Criteria} & \textbf{Step} \\ \hline
\textbf{I$_1$} & Papers should specifically use LLM or Generative AI for Software architecture*  & All \\ \hline
\textbf{E$_1$} & Not in English & T/A \\ 
\textbf{E$_2$} &Duplicated / extension has been included&T/A \\
\textbf{E$_3$} &Out of topic& All \\
\textbf{E$_4$} &Non peer-reviewed papers & T/A \\
\textbf{E$_5$} &Not accessible by institution& T/A\\
\textbf{E$_6$} & Papers mentioning software architecture for running LLM or Gen-ai& F\\
\textbf{E$_7$} &Papers before 15.3.2022 when the initial release of GPT-3.5 was made publicly available** & F \\\hline
\multicolumn{3}{l}{*The papers should genuinely be talking about LLM and SA, } \\
\multicolumn{3}{l}{not just mentioning the buzzword in abstracts/discussion}\\
\multicolumn{3}{l}{**https://platform.openai.com/docs/models}

\end{tabular}
\end{table}

We only included a paper that specifically uses LLM or GenAI for Software architecture (T/A), defines these terms (F), reports causes or factors of this phenomenon (F), proposes approaches or tools for their measurement (F), and recommends any techniques or approaches for remediation (F).
 
 In the exclusion criteria, we excluded a paper that was not written in English (T/A), was duplicated, or had an extension already included in the review (T/A), they were beyond the scope (All), or was not accessible by an institution (T/A).

\subsubsection{Search and Selection Process for the Peer-Reviewed Papers (white)} 
We conducted the search and selection process in February 2025 and included all available publications until this period. The application of the search terms returned \textbf{621 unique white papers} as reported in Table~\ref{tab:SelectionResults}. 

\begin{itemize}
\item \textit{Testing the applicability of the inclusion and exclusion criteria:} Before implementing the inclusion and exclusion criteria, we evaluated their applicability~\cite{Kitchenham2013} in ten randomly chosen articles from the retrieved paper (assigned to all authors). 

\item \textit{Applying inclusion and exclusion criteria to the title and abstract:} We used the same criteria for the remaining 611 articles. Two authors read each paper, and if there was any disagreement, a third author participated to resolve the disagreement. 
We included a third author for 30 papers. The interrater agreement through the Cohen coefficient $k$ showed a 71\% agreement corresponding to a substantial agreement. 
Based on the title and abstract, we selected 45 of the original 621 papers.

\item \textit{Full reading:} 
We performed a full read of the 45 papers included by title and abstract, applying the inclusion and exclusion criteria defined in Table~\ref{tab:Criteria} and assigning each article to two authors. We involved a third author for eight papers to reach a final decision. Based on this step, we selected \ReviewerA{27} papers as possibly relevant contributions (Cohen's $k$ coefficient 64\%: substantial agreement). 

\item \textit{Snowballing:} The snowballing process~\cite{Wohlin2014} involved: 1) the evaluation of all articles that cited the recovered articles and 2) the consideration of all references in the recovered articles. The snowball search was performed in February 2025. We found that \ReviewerA{11} articles were included in the final set of publications.
Since our search and selection process was conducted immediately after the notification of the International Conference on Software Architecture (ICSA) 2025, we waited for the pre-print of all accepted papers to be available to avoid not including some potentially interesting contributions.

\item \textit{Quality and Assessment Criteria:} Before proceeding with the review, we checked whether the quality of the selected articles was sufficient to support our goal and whether the quality of each article reached a certain quality level. We perform this step according to the protocol proposed by Dyb{\aa} and Dings{\o}yr~\cite{Dyba2008}. 
To evaluate the selected articles, we prepared a checklist (Table~\ref{tab:QualityCriteriaWhite}) with a set of specific questions. We rank each answer, assigning a score on a five-point Likert scale (0=poor, 4=excellent). A paper satisfied the quality assessment criteria if it achieved a rating higher than (or equal to) 2.
Among the 39 papers included in the review of the search and selection process, only 37 fulfilled the quality assessment criteria, as reported in Table~\ref{tab:SelectionResults}.
\end{itemize} 

Starting from the \ReviewerA{621} unique papers, following the process, we finally included \textbf{36} papers as reported in Table~\ref{tab:SelectionResults}. 

\begin{table}
\centering
\caption{Quality Assessment Criteria - Peer-Reviewed Papers (white)} 
\label{tab:QualityCriteriaWhite} 
\resizebox{\linewidth}{!}{%
\begin{tabular}{p{0.8cm}|p{7.5cm}}
\hline
\textbf{QA$_s$} & \textbf{QA} \\ \hline
QA$_1$&Is the paper based on research (or is it merely a ``lessons learned'' report based on expert opinion)? \\ \hline 
QA$_2$&Is there a clear statement of the aims of the research? \\\hline 
QA$_3$&Is there an adequate description of the context in which the research was carried out?  \\\hline 
QA$_4$&Was the research design appropriate to address the aims of the research? \\\hline 
QA$_5$ & Was the recruitment strategy appropriate for the aims of the research?  \\\hline 
QA$_6$&Was there a control group with which to compare treatments?\\ \hline 
QA$_7$&Was the data collected in a way that addressed the research issue? \\\hline 
QA$_8$&Was the data analysis sufficiently rigorous? \\\cline{1-2}
QA$_9$&Has the relationship between researcher and participants been considered to an adequate degree?\\\cline{1-2}
QA$_{10}$&Is there a clear statement of findings?\\\cline{1-2}
QA$_{11}$&Is the study of value for research or practice?\\\hline
\multicolumn{2}{l}{\small{Response scale: 4 (Excellent), 3 (Very Good), 2 (Good),}} \\
\multicolumn{2}{l}{\small{1 (Fair), 0 (Poor)}} 
\end{tabular}%
}
\end{table}

\begin{table*}
    \centering
    \caption{Quality Assessment Criteria - Grey literature}
    \label{tab:QualityCriteria}
    \resizebox{\linewidth}{!}{%
        \begin{tabular}{l|p{9cm}|p{9cm}}
\hline
\textbf{Criteria}    & \textbf{Questions} & \textbf{Possible Answers}\\ \hline 
Authority of the producer & Is the publishing organization reputable?	& 1: reputable and well known organization 	\\ \cline{3-3} 
& & 0.5: existing organization but not well known, 0: unknown or low-reputation organization	\\ \cline{2-3}
& Is an individual author associated with a reputable organization?	& 1: true	\\ \cline{3-3}
& & 0: false	\\ \cline{2-3}

& Has the author published other work in the field?	 & 1: Published more than three other work\\  \cline{3-3}
&&0.5: published 1-2 other works, 0: no other works published.	\\  \cline{2-3}

& Does the author have expertise in the area? (e.g., job title principal software engineer)	& 1: author job title is principal software engineer, cloud engineer, front-end developer or similar\\  \cline{3-3} 
&& 0: author job not related to any of the previously mentioned groups. ) \\ \hline 

Methodology & Does the source have a clearly stated aim? & 1: yes\\  \cline{3-3}
&&  0: no	\\ \cline{2-3}

& Is the source supported by authoritative, documented references?	& 1: references pointing to reputable sources \\\cline{3-3}
&&0.5: references to non-highly reputable sources\\  \cline{3-3}
&& 0: no references\\  \cline{2-3}

& Does the work cover a specific question?	& 1: yes\\ \cline{3-3}
&& 0.5: not explicitly\\ \cline{3-3}
&& 0: no\\ \hline 

Objectivity & Does the work seem to be balanced in presentation	& 1: yes\\ \cline{3-3}
&& 0.5: partially\\ \cline{3-3}
&& 0: no\\  \cline{2-3}

& Is the statement in the sources as objective as possible? Or, is the statement a subjective opinion?	 & 1: objective\\ \cline{3-3}
&& 0.5 partially objective\\ \cline{3-3}
&& 0: subjective\\  \cline{2-3}

& Are the conclusions free of bias or is there vested interest? E.g., a tool comparison by authors that are working for a particular tool vendor & 1=no interest\\ \cline{3-3}
&& 0.5: partial or small interest\\ \cline{3-3}
&& 0: strong interest\\  \cline{2-3}

& Are the conclusions supported by the data? & 1: yes \\ \cline{3-3}
&& 0.5: partially\\ \cline{3-3}
&& 0: no\\ \hline 

Date & Does the item have a clearly stated date? & 1: yes\\ \cline{3-3}
&& 0: no \\  \hline 

Position w.r.t. related sources & Have key related GL or formal sources been linked to/discussed? & 1: yes\\ \cline{3-3}
&& 0: no \\\hline 

Novelty & Does it enrich or add something unique to the research?	& 1: yes\\ \cline{3-3}
&& 0.5: partially\\ \cline{3-3}
&& 0: no \\\hline 

Outlet type & Outlet Control & 1:  high outlet control/ high credibility: books, magazines, theses, government reports, white papers \\ \cline{3-3}
& & moderate outlet control/ moderate credibility: annual reports, news articles, videos, Q/A sites (such as StackOverflow), wiki articles \\ \cline{3-3}
& &  0: low outlet control/low credibility: blog posts, presentations, emails, tweets \\ \hline 

\end{tabular}%
}
\end{table*} 

\subsubsection{Search and Selection Process for the Grey Literature} The search was carried out in \ReviewerC{February} 2025 and included all publications available until this period. The application of the search terms returned \textbf{433 unique contributions to the grey literature} as reported in Table~\ref{tab:SelectionResults}.  

\begin{itemize}
\item \textit{Testing the applicability of inclusion and exclusion criteria.} We used the same method adopted in the search and selection process for the peer-reviewed papers (10 papers as a test case) 

\item \textit{Applying inclusion and exclusion criteria to title and abstract.} We applied the criteria to the remaining 423 papers. Two authors read each paper, and if there were disagreements, a third author participated in the discussion to resolve them. For 25 articles, we include a third author. Of the 433 initial papers, we included 77 based on title and abstract (Cohen's $k$ coefficient 81\%: almost perfect agreement). 

\item \textit{Full reading.} We fully read the 77 articles included by title and abstract, applying the criteria defined in Table~\ref{tab:Criteria} and assigning each to two authors. We involve a third author for one paper to reach a final decision (Cohen's $k$ coefficient 88\%: almost perfect agreement). Based on this step, we selected five papers as possibly relevant contributions.

\item \textit{Snowballing.} The snowball search was carried out in February 2025. We found that three articles were included in the final set of publications.

\item \textit{Quality and Assessment Criteria.} Different from peer-reviewed literature, grey literature does not go through a formal review process, and therefore, its quality is less controlled. To evaluate the credibility and quality of the sources selected from the grey literature and to decide whether to include a source from the grey literature or not, we applied the quality criteria proposed by Garousi et al.~\cite{MLRguidelines} (Table~\ref{tab:QualityCriteria}), considering the authority of the producer, the methodology applied, objectivity, date, novelty, impact, and outlet control.
Two authors assessed each source using the aforementioned criteria, with a binary or 3-point Likert scale, depending on the criteria themselves. In case of disagreement, we discuss the evaluation with the third author, who helped provide the final assessment.
We finally calculated the average of the scores and rejected sources from the grey literature that scored less than 0.5 on a scale ranging from 0~to~1.
\end{itemize}

Starting from the \ReviewerA{433} unique papers, following the process, we finally included \textbf{10} grey literature papers as reported in Table~\ref{tab:SelectionResults}. 

\begin{table}
\centering
\footnotesize
\caption{Search and Selection Process} 
\label{tab:SelectionResults} 
\resizebox{0.9\linewidth}{!}{%
\begin{tabular}
{@{}p{7cm}|r@{}}
\hline
\textbf{Step} & \textbf{\#}  \\ \hline
\textbf{Retrieval from white sources }(unique papers) & \textbf{621}   \\ \hline
-Reading by title and abstract& -576 \\
-Full reading  & - 18\\ 
-Snowballing & +  11\\ 
-Quality assessment &  - 2\\ \hline
\textbf{Primary studies} & \textbf{36} \\ \hline \hline 
\textbf{Retrieval from grey sources} (unique papers) & \textbf{433}   \\ \hline
-Reading by title and abstract& -356\\ 
-Full reading  & - 70\\ 
-Snowballing & + 3 \\ \hline
\textbf{Primary studies} & \textbf{10} \\ \hline
\end{tabular}
}
\end{table}
\subsection{Data Extraction}

Starting from the initial \textbf{1054 unique papers} (621 white and 443 grey ), following the process, we finally included \textbf{46 papers} (36 white and 10 grey) as reported in Table~\ref{tab:SelectionResults}.
The data extraction form, together with the mapping of the information needed to answer each RQ, is summarized in Table~\ref{tab:DataExtraction}.
We extracted the data following the open coding approach~\cite{CorbinStrauss2008}, in which two authors extracted the information, and we involved a third author in case of disagreement. This data is exclusively based on what is reported in the papers, without any kind of personal interpretation. 

\begin{table}
\centering
\footnotesize
\caption{Data Extraction}
\label{tab:DataExtraction}
\resizebox{\linewidth}{!}{%
\begin{tabular}{m{2.6cm}|m{0.7cm}|p{4cm}} \hline 
\textbf{Data} & \textbf{RQ}  & \textbf{Outcome} \\ \hline 
Work category & \multirow{9}{*}{na} & List of Category \\ \cline{1-1}\cline{3-3}
Methods & & List of methodological approaches   \\ \cline{1-1}\cline{3-3}
\multirow{2}{*}{Author} &  & First and last name\\
&& Affiliation \\  \cline{1-1}\cline{3-3}
\multirow{5}{*}{Publication Sources} & &   Peer-reviewed literature (white)\\
&&  Grey literature  \\
&&  Publication name \\
&&  Publication type (e.g., journal) \\
&&  Publication year  \\ \cline{1-3}\cline{3-3}

\multirow{3}{*}{GenAI usage}  & RQ$_{1.1}$ & Purpose (why)  \\ \cline{2-3}
& RQ$_{1.2}$ & Model (what) \\\cline{2-3}
& RQ$_{1.3}$ & How \\ \cline{1-1}\cline{2-3}

\multirow{7}{*}{GenAI usage context} & RQ$_{2.1}$ & SALC phase   \\ \cline{2-3}
& RQ$_{2.2}$ & For what architectural styles or patterns  \\ \cline{2-3}
& RQ$_{2.3}$ & For what architectural maintenance / quality-related tasks \\ \cline{2-3}
& RQ$_{2.4}$ &  Architecture analysis / modeling method \\ \cline{1-3}
\multirow{3}{*}{Use case} & \multirow{3}{*}{RQ$_{2.5}$}	& List of use cases  \\
&& Analyzed systems \\
&& Programming languages \\\cline{1-1}\cline{2-3}
Future Challenges & RQ$_3$	& List of challenges \\ \hline
\end{tabular}
}
\end{table}

\section{Results}
\label{sec:results}

% \todo[inline]{Missing RQ-Result-TakeAway title harmonization}
In this Section, we report the results to answer our RQs. \ReviewerB{From this section onward, we visually and textually distinguish results from white literature using \textcolor{whiteLitteratureColor}{\textbf{\whiteLitteraturePrefix}}, and from gray literature using \textcolor{grayLitteratureColor}{\textbf{\grayLitteraturePrefix}}.}

\subsection{Study Context}
This sub-section provides an overview of the study context in the reviewed research, including the types of studies conducted, the balance between white and gray literature, and the categories of published works. 
\begin{table}
\centering
\footnotesize
\caption{White and Grey Literature Distribution}
\label{tab:study_context:wglitterature}
\footnotesize
\resizebox{\linewidth}{!}{%
\begin{tabular}{l p{5.2cm} rr}
\hline
\textbf{Code }  & \textbf{PaperID}                                                                & \textbf{\#}     & \textbf{\%} \\ \hline
White &\wlcite{Adnan25,Ahmad23,Arias24, Arun25, Dhar24a,Dhar24b,Diaz24,Diaz25,Duarte25,Eisenreich24,Fuchb25,Hagel25,Heiben24,Ivers25,Jahic24,Johansson24,Jose24,Lutze24,Manuel24,Mino24,Pandini25,Quevedo24,Raghavan24,Rejithkumar24,Ronau24,Rubei25,Rukmono23,Rukmono24,Saarinen24,Schindler24,Singh24,Soliman25,Supekar24,Tagliaferro25,Tang23,Wei24} & 36& 78\%                   \\ 

Grey  &\glcite{Ahuja24,Chandraraj23,Data23,Fujitsu25,Sharma24,Martelli23,Nandi24,Paradkar23,Prakash24,Seroter23}& 10 & 22\%                   \\\hline
\end{tabular}%
}
\end{table}
% \usepackage{graphicx}
\begin{table}
\centering
\footnotesize
\caption{Study Type}
\label{tab:study_context:WorkType}
\resizebox{\linewidth}{!}{%
\begin{tabular}{l p{4cm} r r}
\hline
\textbf{Code }                                                                                                                                & \textbf{PaperID}                                                                     & \textbf{\#}& \textbf{\%}\\ \hline
Case Study        & \wlcite{Ahmad23,Arias24,Dhar24a,Dhar24b,Diaz24,Diaz25,Eisenreich24,Jahic24,Johansson24,Jose24,Lutze24,Pandini25,Raghavan24,Rejithkumar24,Ronau24,Rubei25,Rukmono23,Rukmono24,Saarinen24,Sharma24,Schindler24,Singh24,Soliman25,Tagliaferro25,Tang23,Wei24},\glcite{Prakash24,Manuel24} & 28    &  40\%             \\
Experiment        & \wlcite{Adnan25,Arun25,Duarte25,Fuchb25,Hagel25,Mino24,Quevedo24,Ronau24,Soliman25}, \glcite{Ahuja24}                                                                                                                                                                                                             & 10 & 14\%                 \\
Exploratory Study & \wlcite{Ivers25,Arun25}                                                                                                                                                                                                                                                                                                                   & 2    &  3\%             \\
Method Proposal   & \wlcite{Adnan25,Dhar24b,Diaz24,Diaz25,Duarte25,Eisenreich24,Fuchb25,Hagel25,Heiben24,Jose24,Lutze24,Pandini25,Rubei25,Rukmono24,Supekar24,Tagliaferro25,Wei24}, \glcite{Ahuja24,Prakash24,Manuel24}                                                                                 & 20      & 29\%           \\
PoC               & \wlcite{Heiben24,Rukmono24}                                                                                                                                                                                                                                                                                                     & 2    & 3\%              \\
Survey            & \wlcite{Jahic24}                                                                                                                                                                                                                                                                                                                   & 1  & 1\%                \\
Tool Review       & \glcite{Chandraraj23,Data23,Fujitsu25,Martelli23,Nandi24,Paradkar23,Seroter23}                                                                                                                                                                                                                                   & 7         & 10\%        \\\hline
\end{tabular}%
}
\end{table}

\begin{table}
\centering
\caption{Study Category}
\label{tab:study_context:work_category}
\footnotesize
\resizebox{\linewidth}{!}{%
\begin{tabular}{l p{4.5cm} rr }
\hline
\textbf{Code }                                                                                                                                & \textbf{PaperID}             & \textbf{\#}                                                        & \textbf{\%} \\ \hline
Blog Post       & \glcite{Chandraraj23,Nandi24,Paradkar23,Seroter23}                                                                                                                                                                                                                              &  4    & 9\%                    \\
Full Paper      & \wlcite{Adnan25,Arias24,Arun25, Dhar24a,Diaz24,Diaz25,Duarte25,Fuchb25,Hagel25,Heiben24,Ivers25,Johansson24,Jose24,Lutze24,Pandini25,Quevedo24,Ronau24,Rubei25,Saarinen24,Schindler24,Soliman25,Supekar24,Tagliaferro25,Wei24}, \glcite{Ahuja24} & 25 & 54\%                   \\
Industry Report & \wlcite{Singh24}                                                                                                                                                                                                                                                                              &1 & 2\%                    \\
Position Paper  & \wlcite{Sharma24}                                                                                                                                                                                                                                                                             &1 & 2\%                    \\
Short Paper     & \wlcite{Ahmad23,Jahic24,Mino24,Raghavan24,Rejithkumar24,Rukmono23,Tang23}                                                                                                                                                                                                   & 7 & 15\%                  \\
Thesis          & \glcite{Manuel24,Prakash24}                                                                                                                                                                                                                                    & 2     &  4\%                  \\
Vision Paper    & \wlcite{Dhar24b,Eisenreich24,Rukmono24}                                                                                                                                                                                                                                                    & 3     & 7 \%                 \\
White Paper     & \glcite{Fujitsu25,Martelli23}                                                                                                                                                                                                                                                           & 2   & 4\%                    \\
Youtube Video   & \glcite{Data23}                                                                                                                                                                                                                                                                     &1          & 2\%                   \\ \hline
\end{tabular}%
}
\end{table}
\begin{table*}[t]
\centering
\footnotesize
\caption{Publication Sources}
\label{tab:study_context:publication_venue_name}
\resizebox{\linewidth}{!}{%
    \begin{tabular}{p{12cm} l rr}
        \hline
        \textbf{Sources Name} & \textbf{Type} & \textbf{Count} & \textbf{Years} \\ \hline
        AIM Research & Research Institution & 1 & - \\ 
        Communications in Computer and Information Science & Book Series & 1 & - \\ 
        Design Society & Society Publication & 1 & - \\ 
        Electronics (Switzerland) & Journal & 1 & - \\ 
        European Conference on Pattern Languages of Programs, People and Practices & Proceedings & 1 & - \\ 
        European Conference on Software Architecture & Conference & 1 & 2024 \\ 
        Human-Computer Interaction & Journal & 1 & - \\ 
        IEEE International Conference on Software Quality Reliability and Security Companion (QRS-C) & Proceedings & 1 & 2023 \\ 
        IEEE International Conference on Data and Software Engineering (ICoDSE) & Proceedings & 1 & 2023 \\
        IEEE International Requirements Engineering Conference (RE) & Conference & 1 & 2024 \\ 
        IEEE International Conference on Software Architecture (ICSA) & Conference & 12 & 2024, 2025 \\ 
        IEEE International Conference on Software Architecture Companion (ICSA-C) & Conference & 3 & 2024 \\ 
        IEEE Software & Journal & 1 & - \\ 
        IEEE/ACM Workshop on Multi-disciplinary Open and RElevant Requirements Engineering (MO2RE) & Workshop & 1 & 2024 \\ 
        Information Technology & Journal & 1 & - \\ 
        Institutional Website & Website & 7 & - \\
        International Conference on Software Engineering & Proceedings & 1 & - \\
        International Workshop on Designing Software & Workshop & 1 & 2024 \\ 
        Lecture Notes in Computer Science (including subseries Lecture Notes in Artificial Intelligence and Lecture Notes in Bioinformatics) & Book Series & 1 & - \\
        Medium & Online Media & 2 & - \\
        Methods & Journal & 1 & - \\ 
        SN Computer Science & Journal & 1 & - \\ 
        Studies in Computational Intelligence & Book Series & 1 & - \\ 
        YouTube & Online Media & 1 & - \\ \hline
    \end{tabular}
}
\end{table*}

\begin{figure}
    \centering
    \includegraphics[width=\linewidth]{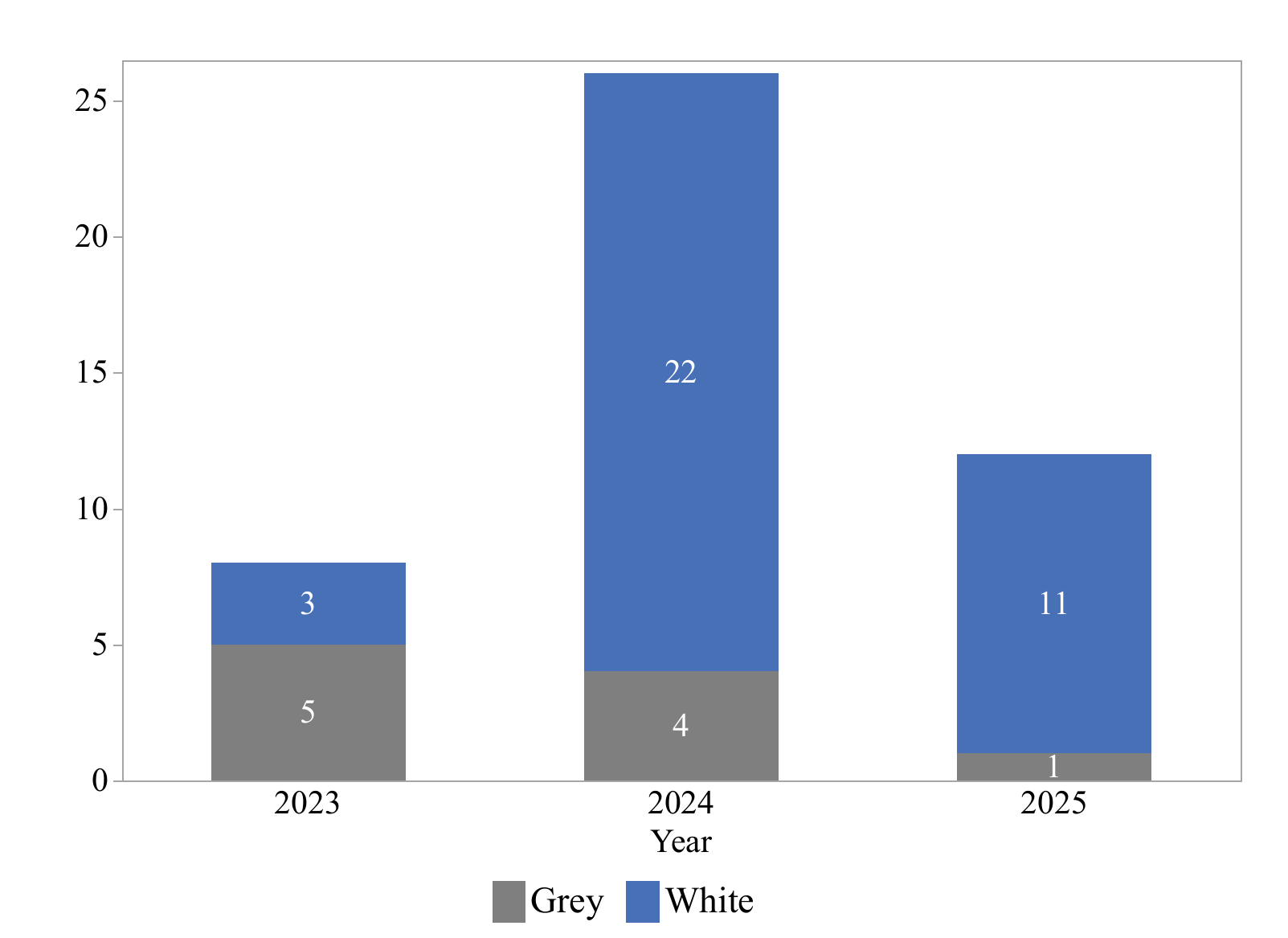}
    \caption{Publication Source Trend}
    \label{fig:publicationtrend}
\end{figure}

Most of the works we considered belong to white literature (36; 78\%) while 22\% (10) to the gray (Table \ref{tab:study_context:wglitterature}).
Case studies are the most common type (28; 40\%), followed by method proposals (20, 29\%) and experiments (10; 14\%). Tool reviews are proposed only from gray literature (7; 10\%) while proof-of-concept (PoC) studies (2; 3\%) interestingly are represented only by white literature. Surprisingly, we only included a few position papers (1; 2\%) \wlcite{Sharma24} and vision papers (3; 7\%) (Table \ref{tab:study_context:WorkType}). Most of them are full papers (25; 52\%), followed by short papers (7; 15\%) and a few Thesis (2; 4\%) (Table \ref{tab:study_context:work_category}).
Finally, according to Figure \ref{fig:publicationtrend}, GenAI in SA was prominently discussed and featured in the gray literature during the start of the hype (2023), but the white literature became prominent the year after consolidating in 2025 as the main publication source for the topic.

\subsection{Generative AI for Software Architecture: How is it used (RQ$_1$)}
Here, we present how GenAI is currently applied in SA in terms of purpose, models used, and techniques for performance improvement, such as prompt engineering practices and the level of human interaction.

\subsubsection{Why GenAI in SA (RQ$_{1.1}$)}
Architectural decision support is the purpose most frequently investigated in the reviewed studies, appearing in 38\% (18) of them (Table \ref{tab:RQ1:purpose_LLM} - \textbf{RQ$_{1.1}$}). This suggests that the primary focus of current research on GenAI in software architecture is its application in assisting architectural decision-making. For example, \wlcite{Arias24} uses GenAI to generate microservice names, while~\glcite{Prakash24} uses it to support software design and requirement engineering, and~\wlcite{Jose24} uses it to guide software architects in making architectural decisions. Similarly, the second most frequent purpose for using GenAI in the case of reverse engineering for architectural reconstruction appears in 19\% (9) of the cases. On the other hand, the least explored uses are Reverse Engineering for Traceability (\glcite{Fuchb25}) and Migration \& Re-engineering (\wlcite{Singh24}), each of which appeared only in 2\% (1) of the studies (Table \ref{tab:RQ1:purpose_LLM} - \textbf{RQ$_{1.1}$}).

\begin{table*}[t]
\centering
\caption{Purpose of the LLM - (RQ$_{1.1}$)}
\label{tab:RQ1:purpose_LLM}
\footnotesize
\resizebox{\linewidth}{!}{%
\begin{tabular}{m{7cm} p{6cm} rr}
\hline
                                            \textbf{Code} &                                            \textbf{PaperID} &  \textbf{Count*} &  \textbf{\%} \\
\hline
                  Architectural Decision Support & \wlcite{Ahmad23, Arias24, Dhar24a, Diaz24, Diaz25, Hagel25, Jahic24, Jose24, Quevedo24, Raghavan24, Supekar24}, \glcite{Chandraraj23, Data23, Manuel24,Martelli23, Nandi24, Prakash24,Paradkar23} &     18 & 38\% \\
Reverse Engineering/Architectural Reconstruction & \wlcite{Duarte25, Johansson24, Mino24, Pandini25, Rubei25, Rukmono24, Schindler24}, \glcite{Fujitsu25, Seroter23} &     9 & 19\% \\
                         Architecture Generation & \wlcite{Adnan25, Dhar24b, Eisenreich24, Heiben24, Lutze24, Saarinen24, Sharma24,Tagliaferro25, Wei24} &      9 & 19\% \\
                              Quality Assessment &       \wlcite{Duarte25, Mino24, Pandini25, Rubei25} &      4 &  9\% \\
                          Software Comprehension &              \wlcite{Ronau24, Rukmono23, Soliman25} &      3 &  7\% \\
                         Requirement Engineering &                      \wlcite{Rejithkumar24, Tang23} &      2 &  4\% \\
                      Migration \& Re-engineering &                                    \wlcite{Singh24} &      1 &  2\% \\
                Reverse Engineering/Traceability &                                    \wlcite{Fuchb25} &      1 &  2\% \\
\hline
*One paper can have more than one purpose \\
\end{tabular}
}
\end{table*}

\begin{keyTakeAways}[RQ$_{1.1}$ (Why GenAI in SA)]
LLMs are primarily used for architectural decision support (38\%) and reverse engineering (21\%), with less focus on tasks like migration, re-engineering, and traceability.
\end{keyTakeAways}

\subsubsection{GenAI Model Used (RQ$_{1.2}$)}
OpenAI GPT models are the ones that rule the roost and were utilized in 62\% (105) of the articles, followed by Google's models (15; 9\%)  (Table \ref{tab:RQ1:LLMModelsMerged} - \textbf{RQ$_{1.2}$}). Surprisingly, the recently published open-source model DeepSeek has already been applied in two works. It is also worth noting that on-demand cloud-based models are by far the favorable option in place of on-premises due to their resource requirements.

\begin{figure}
    \centering
    \includegraphics[width=\linewidth]{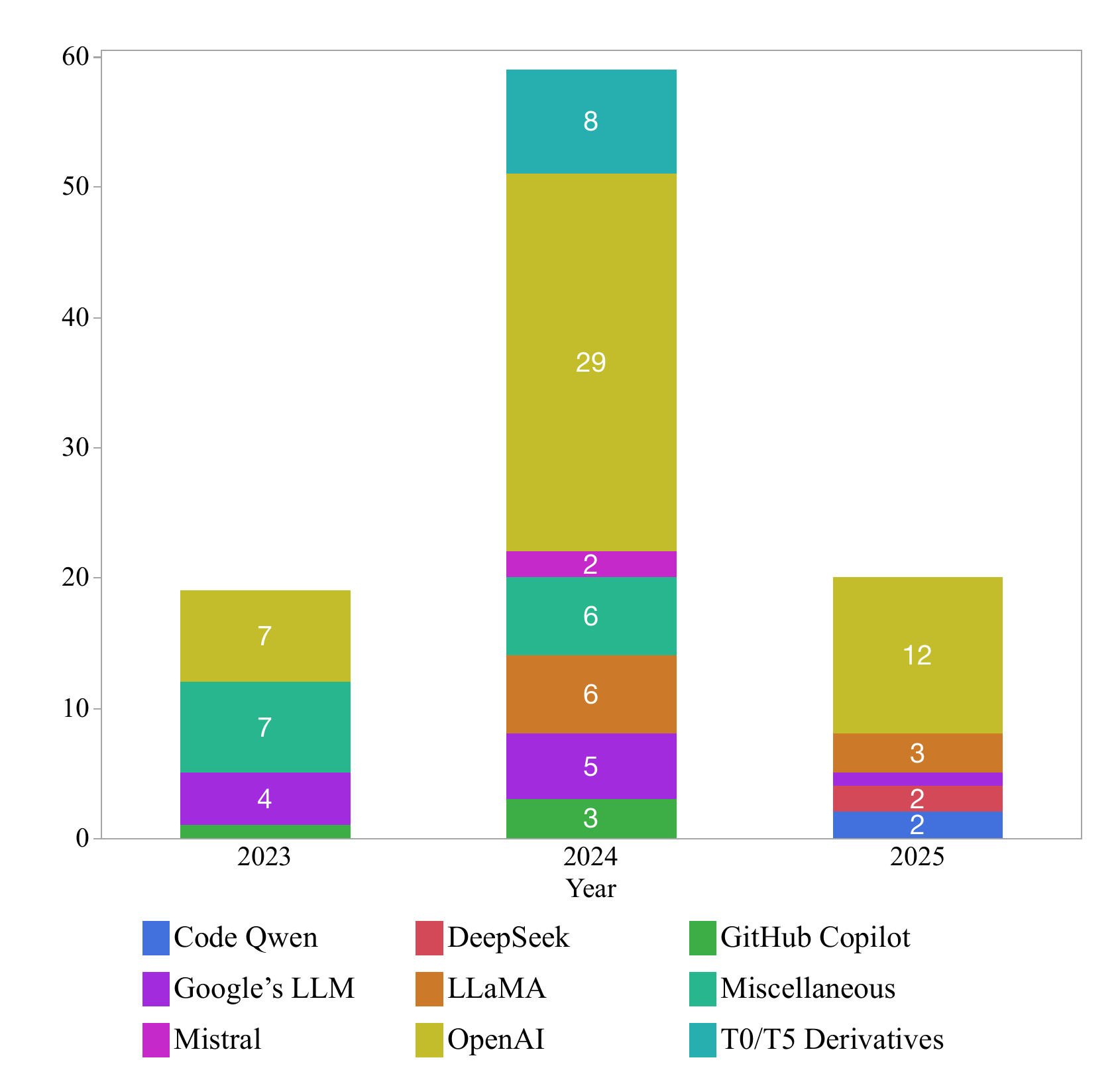}
    \caption{LLM Vendor Trend (RQ$_{1.2}$) }
    \label{fig:LLMVendorTrend}
\end{figure}

\ReviewerB{
Considering the evolution over time of the AI model providers, it is evident that OpenAI has a consistent prominence. Nevertheless, newer models such as DeepSeek and Qwen gained traction in 2024 and 2025, highlighting a shift in attention toward emerging alternatives. This trend is also confirmed in the increasing presence of models from diversified providers, i.e., miscellaneous category, including specific open source alternatives such as LLaMa (Figure\ref{fig:LLMVendorTrend}  - \textbf{RQ$_{1.2}$}).
}

\begin{table*}
\caption{LLM Models  - (RQ$_{1.2}$)}
\label{tab:RQ1:LLMModelsMerged}
\centering
\resizebox{0.9\linewidth}{!}{%
\footnotesize

\begin{tabular}{llp{8.5cm}rrr}
\hline
\textbf{Model Family }                      &\textbf{ Model }& \textbf{PaperID }                                                                                                                                                                                                                                                                                                                                                                                       & \textbf{Count* }& \textbf{\% (Model)} & \textbf{\% (Family)}            \\ \hline
\multirow{16}{*}{OpenAI}           & GPT                      & \wlcite{Adnan25, Ahmad23, Arias24, Arun25, Dhar24a, Dhar24b, Diaz24, Diaz25, Duarte25, Eisenreich24, Fuchb25, Hagel25, Heiben24, Ivers25, Jahic24, Johansson24, Jose24, Lutze24, Mino24, Pandini25, Prakash24, Quevedo24, Rejithkumar24, Ronau24, Rubei25, Rukmono23, Rukmono24, Saarinen24, Schindler24, Singh24, Soliman25, Tagliaferro25, Tang23, Wei24}, \glcite{ Chandraraj23, Data23, Manuel24, Martelli23, Nandi24, Paradkar23, Seroter23} & 39 & 23\% & \multirow{16}{*}{62\% } \\
                                   & GPT-4                    & \wlcite{Adnan25, Dhar24a, Dhar24b, Diaz25, Duarte25, Fuchb25, Hagel25, Heiben24, Ivers25, Jahic24, Jose24, Lutze24, Pandini25,  Ronau24, Rubei25, Rukmono24, Schindler24, Singh24, Tagliaferro25} \glcite{Prakash24} & 21 & 13\% &                       \\
                                   & ChatGPT                  & \wlcite{Ahmad23, Arias24,Arun25,   Mino24, Quevedo24, Saarinen24, Tang23, Wei24}, \glcite{Chandraraj23, Data23, Manuel24, Martelli23, Paradkar23} & 14 & 8\% &                       \\
                                   & GPT-3                    & \wlcite{Dhar24a, Dhar24b, Diaz24, Jahic24, Johansson24, Rejithkumar24, Rukmono23, Schindler24, Soliman25} & 9 & 5\% &                       \\
                                   & GPT-3.5                  & \wlcite{Dhar24a, Dhar24b, Diaz24, Johansson24, Rejithkumar24, Rukmono23, Schindler24, Soliman25} & 8 & 5\% &                       \\
                                   & GPT-4o                   & \wlcite{Adnan25, Diaz25, Duarte25, Fuchb25, Hagel25, Rubei25, Tagliaferro25} & 7 & 4\% &                       \\
                                   & GPT-4o-mini              & \wlcite{Adnan25, Diaz25, Duarte25} & 3 & 2\% &                       \\
                                   & GPT-2                    & \wlcite{Dhar24a, Dhar24b} & 2 & 1\% &                       \\
                                   & GPT-3.4                  & \wlcite{Jahic24} & 1 & 1\% &                       \\
                                   & GPT-4 Turbo              & \wlcite{Ronau24} & 1 & 1\% &                       \\
\hline \multirow{6}{*}{Google's LLM}      & Bard                     & \wlcite{Lutze24, Mino24, Schindler24}, \glcite{Data23, Martelli23, Seroter23} & 6 & 4\% & \multirow{6}{*}{9\% }   \\
                                   & Gemini                   & \wlcite{Rubei25, Saarinen24}, \glcite{Seroter23} & 3 & 2\% &                       \\
                                   & Google Bard              & \wlcite{Mino24, Schindler24}, \glcite{Martelli23} & 3 & 2\% &                       \\
                                   & Bert                     & \wlcite{Dhar24b} & 1 & 1\% &                       \\
                                   & Gemini 1.5               & \wlcite{Rubei25} & 1 & 1\% &                       \\
                                   & Google Gemini            & \wlcite{Saarinen24} & 1 & 1\% &                       \\
\hline \multirow{6}{*}{LLaMA}             & LLaMA                    & \wlcite{Eisenreich24, Fuchb25, Heiben24, Pandini25, Sharma24}, \glcite{Ahuja24, Manuel24} & 7 & 4\% & \multirow{6}{*}{8\% }   \\
                                   & LLaMA-3                  & \glcite{Manuel24} & 2 & 1\% &                       \\
                                   & Llama 3.1                & \wlcite{Fuchb25} & 1 & 1\% &                       \\
                                   & LLaMA-2                  & \glcite{Ahuja24} & 1 & 1\% &                       \\
                                   & Code Llama               & \glcite{Sharma24} & 1 & 1\% &                       \\
                                   & Codellama 13b            & \wlcite{Fuchb25} & 1 & 1\% &                       \\
\hline \multirow{2}{*}{DeepSeek}          & DeepSeek-Coder           & \wlcite{Arun25} & 1 & 1\% & \multirow{2}{*}{1\% }   \\
                                   & DeepSeek-V2.5            & \wlcite{Adnan25} & 1 & 1\% &                       \\
\hline \multirow{2}{*}{CodeQwen}          & CodeQwen                 & \wlcite{Adnan25} , \wlcite{Arun25} & 2 & 1\% & \multirow{2}{*}{2\% }   \\
                                   & CodeQwen1.5-7B           & \wlcite{Adnan25} & 1 & 1\% &                       \\
\hline GitHub Copilot                     & Copilot                  & \wlcite{ Saarinen24}, \glcite{Prakash24,Martelli23, Nandi24} & 4 & 2\% & 2\%                     \\
\hline \multirow{2}{*}{Mistral}           & Mistral                  & \glcite{Ahuja24, Rejithkumar24} & 2 & 1\% & \multirow{2}{*}{2\% }   \\
                                   & Mistral 7b               & \wlcite{Rejithkumar24} & 1 & 1\% &                       \\
\hline \multirow{5}{*}{T0/T5 Derivatives} & T5                       & \wlcite{Dhar24a, Dhar24b}, \wlcite{Sharma24} & 3 & 2\% & \multirow{5}{*}{6\% }   \\
                                   & Flan-T5                  & \wlcite{Dhar24a, Dhar24b} & 2 & 1\% &                       \\
                                   & T0                       & \wlcite{Dhar24a, Dhar24b} & 2 & 1\% &                       \\
                                   & CodeT5                   & \wlcite{Sharma24} & 1 & 1\% &                       \\
                                   & CodeWhisperer            & \glcite{Prakash24} & 1 & 1\% &                       \\
\hline \multirow{15}{*}{Miscellaneous}    & Adobe Firefly            & \glcite{Nandi24} & 1 & 1\% & \multirow{15}{*}{1\% }  \\
                                   & Claude AI                & \wlcite{Singh24} & 1 & 1\% &                       \\
                                   & Codex                    & \wlcite{Sharma24} & 1 & 1\% &                       \\
                                   & Codium                   & \glcite{Martelli23} & 1 & 1\% &                       \\
                                   & Cursor                   & \glcite{Martelli23} & 1 & 1\% &                       \\
                                   & Falcon                   & \wlcite{Eisenreich24} & 1 & 1\% &                       \\
                                   & k8sgpt                   & \glcite{Martelli23} & 1 & 1\% &                       \\
                                   & Mutable.AI               & \glcite{Martelli23} & 1 & 1\% &                       \\
                                   & N.A                      & \wlcite{Raghavan24}, \glcite{Fujitsu25} & 2 & 1\% &                       \\
                                   & Phi-3                    & \glcite{Ahuja24} & 1 & 1\% &                       \\
                                   & Replit                   & \glcite{Martelli23} & 1 & 1\% &                       \\
                                   & Robusta ChatGPT bot      & \glcite{Martelli23} & 1 & 1\% &                       \\
                                   & Tabnine                  & \glcite{Martelli23} & 1 & 1\% &                       \\
                                   & Unknown                  & \wlcite{Supekar24} & 1 & 1\% &                       \\
                                   & Yi                       & \wlcite{Eisenreich24} & 1 & 1\% &                       \\ \hline
\multicolumn{2}{l}{*One paper can have more than one model} \\
\end{tabular}%
}
\end{table*}

\begin{keyTakeAways}[RQ$_{1.2}$ (GenAI Model Used)]
OpenAI GPT models dominate (62\%) the research landscape, while alternatives such as Google LLMs and LLaMA models are significantly less employed.
\end{keyTakeAways}

\subsubsection{How GenAI is used (RQ$_{1.3}$)}
Among the techniques to enhance the capabilities and performance of GenAI, Fine-Tuning is applied in 12\% (6) of the studies, that is, some researchers have chosen to fine-tune LLMs for specific architectural tasks with additional training. In particular, \wlcite{Arun25} used Fine-Tuning to align the LLM in generating serverless functions. RAG, including proprietary variants, is applied in 20\% (10)  of the studies, suggesting that applying external knowledge sources is a common method to improve LLM performance in software architecture contexts. For example, \wlcite{Dhar24b} used RAG and Fine-Tuning to retrieve architecture knowledge management information and align such models to their needed task (Table \ref{tab:genAIused} - \textbf{RQ$_{1.3}$}).

A large percentage of studies (13; 25\% for Prompt Engineering and 24; 48\% for Model Enhancements) did not report any data. Conversely, 26\% (12) reported that no improvements were applied, and the models were run as they were. Therefore, our findings reveal that while fine-tuning and RAG methods are explored, most studies do not document their method of improvement or apply the off-the-shelf models without any modifications (Table \ref{tab:genAIused} - \textbf{RQ$_{1.3}$}).

Most specifically, prompt engineering is also used to quickly align LLMs to a new task \cite{esposito_large_2024}. The most widely used technology is the few-shot prompt, present in 31\% (16). This shows that researchers use numerous examples to a great extent to allow LLMs to produce more precise and contextual architectural output. In contrast, one-shot prompting is the least used, with the technique mentioned in only 2\% (1) of the research, suggesting that a single occurrence is infrequent in this field. Zero-shot prompting occurs in 12\% (6) of the studies, at moderate frequency, where the researchers solely utilize the pre-training knowledge of the model without additional context. As an example, \glcite{Martelli23} employed the three techniques to evaluate LLM applications in modernizing the architecture of legacy systems. Finally, in the spectrum of reasoning enhancements, Chain-of-thought (CoT) prompting appears only in 8\% (4) of the cases.~\wlcite{Wei24} employs such a technique when evaluating an LLM-based pipeline from requirements to code.

% Moreover, 23\% of the articles explicitly state that no type of prompt engineering has been used, while 13\% do not provide any information. Furthermore, 12\% of the articles did not indicate whether or not a prompting strategy had been used, so the data set was somewhat unclear. Hence, we can infer that most articles do not use explicit prompting techniques or at least do not report them (Table \ref{tab:genAIused} Figure \ref{fig:genaibubble} - \textbf{RQ$_{1.3}$}).

Most studies involve some form of human interaction with the model (39; 85\%), and this indicates that our community is prone to involve human observation, validation, or supplementation when using LLMs for software architecture purposes. This indicates that fully autonomous AI-driven architectural decisions are not yet prevalent, but human participation is still significant in guiding, validating, or improving LLM-generated results. For example, \wlcite{Diaz24} leverages human interaction by providing a chat-based environment to provide AI-based support to novice architects to refine design decisions.

\ReviewerB{No human interaction has been reported for 15\% (7) of the studies, and the models existed without direct human intervention.} The breakdown shows a high preference for interactive approaches, validating that LLMs in software development are used primarily as auxiliary tools and not as standalone decision-makers (Table \ref{tab:genAIused} - \textbf{RQ$_{1.3}$}).

\begin{keyTakeAways}[RQ$_{1.3}$ (How GenAI is used)]
Few-shot prompting (31\%) is the most common technique, RAG (22\%) is frequently used for model enhancement, and 85\% of the studies involve human interaction, emphasizing the assistive rather than autonomous role of LLM.
\end{keyTakeAways}

\begin{table}
\centering
\caption{How GenAI is used (RQ$_{1.3}$)}
\label{tab:genAIused}
\resizebox{\linewidth}{!}{%
\begin{tabular}{p{1cm}p{2.5cm}p{5cm}rr}
\hline
                                               & \textbf{Code  }           & \textbf{PaperID }                                                                                                                                                                                                                                                                                                                                                                                                         & \multicolumn{1}{l}{\textbf{Count*}} & \multicolumn{1}{l}{\textbf{\%}} \\ \hline
\multirow{12}{*}{ \rotatebox{90}{Prompt Engineering}}                                                 & Few-Shot         & \wlcite{Adnan25, Arun25,Dhar24a, Diaz24, Duarte25, Hagel25, Heiben24, Ivers25, Jose24, Lutze24, Pandini25, Rubei25, Singh24, Tagliaferro25}, \glcite{ Martelli23, Prakash24} & 16                        & 31\%                   \\
                                               & Unspecified             & \wlcite{Eisenreich24, Manuel24, Quevedo24, Ronau24, Saarinen24, Supekar24}, \glcite{Ahuja24,Chandraraj23, Data23, Fujitsu25,Nandi24,Paradkar23,Seroter23} & 13                   & 25\%                   \\
                                               &  None &	\wlcite{Ahmad23, Arias24, Dhar24b, Jahic24, Johansson24, Mino24, Raghavan24, Rejithkumar24, Rukmono23, Rukmono24, Schindler24, Tang23} &	12 &	23\% \\
                                               & Zero-Shot        & \wlcite{Arun25, Dhar24a, Diaz24, Singh24, Soliman25}, \glcite{Martelli23} & 6                         & 12\%                   \\
                                               & Chain-of-Thought & \wlcite{Diaz25, Fuchb25, Wei24}, \glcite{Sharma24} & 4                         & 8\%                    \\

                                               & One-Shot         & \wlcite{Singh24} & 1                         & 2\%                    \\
                                               \hline
\multirow{12}{*}{\rotatebox{90}{Model Enhancements}} 
                                               & Unspecified             & \wlcite{Adnan25, Diaz25, Duarte25, Fuchb25, Hagel25, Heiben24, Ivers25, Jose24, Lutze24, Quevedo24, Rubei25, Saarinen24, Singh24, Sharma24, Schindler24, Soliman25, Supekar24, Tagliaferro25}, \glcite{Chandraraj23, Data23, Nandi24, Paradkar23,Prakash24, Seroter23} & 24                        & 48\%        \\
                                             & RAG              & \wlcite{Dhar24b, Diaz24, Eisenreich24, Pandini25, Rejithkumar24, Ronau24}, \glcite{Ahuja24, Fujitsu25, Manuel24, Martelli23} & 10                        & 20\%                   \\
                                               & None	& \wlcite{Ahmad23, Arias24, Jahic24, Johansson24, Mino24, Raghavan24, Rukmono23, Rukmono24, Tang23} 	& 9	& 18\% \\
                                               & Fine-Tuning      & \wlcite{Arun25,Dhar24a, Dhar24b, Wei24}, \glcite{Ahuja24,  Martelli23} & 6                         & 12\%                   \\
                                               & Proprietary RAG  &  \glcite{Fujitsu25} & 1                         & 2\%                    \\

                                               \hline
\multirow{8}{*}{\rotatebox{90}{Human Model Interaction}}      
& Yes              & \wlcite{Adnan25, Ahmad23, Arun25, Dhar24b, Diaz24, Diaz25, Duarte25, Eisenreich24, Fuchb25, Hagel25, Heiben24, Ivers25, Jahic24, Johansson24, Jose24, Lutze24, Mino24, Pandini25, Quevedo24, Raghavan24, Rejithkumar24, Ronau24, Rubei25, Rukmono24, Saarinen24, Singh24, Soliman25, Supekar24, Tagliaferro25, Tang23, Wei24}, \glcite{Ahuja24,  Chandraraj23, Data23, Nandi24, Manuel24, Paradkar23, Prakash24, Seroter23} & 39                        & 85\%                   \\ 
& No               & \wlcite{Arias24, Dhar24a, Rukmono23, Sharma24, Schindler24}, \glcite{Fujitsu25, Martelli23} & 7                         & 15\%                   \\
\hline
\multicolumn{2}{c}{Model used as-is}                                  &	\wlcite{Ahmad23, Arias24, Dhar24b, Jahic24, Johansson24, Mino24, Raghavan24, Rejithkumar24, Rukmono23, Rukmono24, Schindler24, Tang23} &	12                         & 26\%                   \\ \hline
\multicolumn{4}{l}{*One paper can have more than one usage} \\
\end{tabular}%
}
\end{table}

\begin{figure*}[htb]
    \centering
    \includegraphics[width=0.7\linewidth]{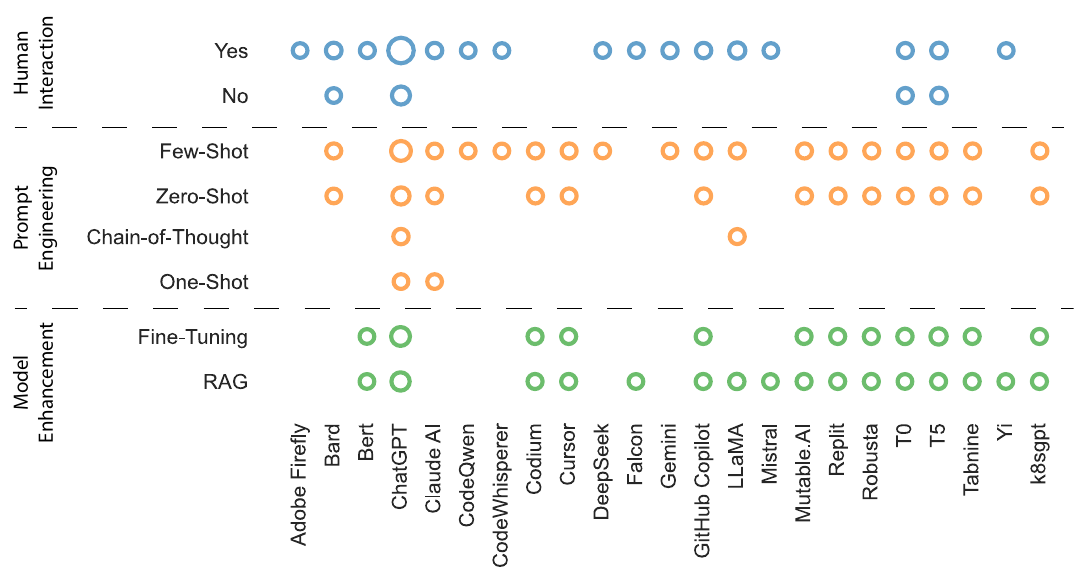}
    \caption{How GenAI is used}
    \label{fig:genaibubble}
\end{figure*}

\subsection{Generative AI for Software Architecture: In which context (RQ$_2$) }
This section presents the different contexts in which GenAI is applied within the software architecture. Specifically, we examine its role across various phases of the Software Architecture Lifecycle (SALC), the architectural styles and patterns it supports, and the validation methods used to assess its outputs.

\begin{figure*}[htb]
    \centering
    \includegraphics[width=\linewidth]{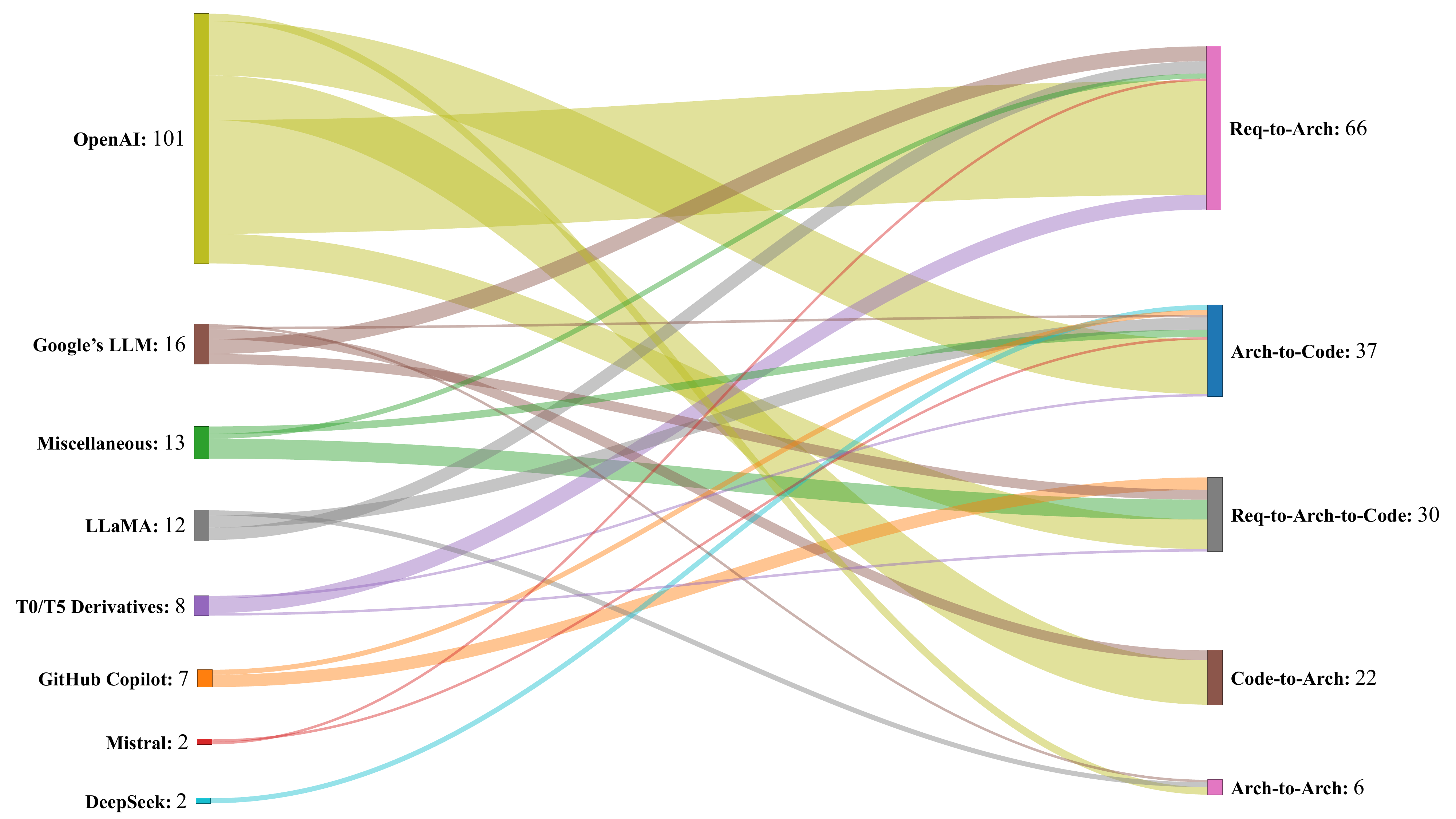}
    \caption{Sankey Plot connecting LLM Models to SALC Phase}
    \label{fig:sankey}
\end{figure*}

\subsubsection{SALC Phases (RQ$_{2.1}$)}
Regarding the use of GenAI across SALC (Table \ref{tab:rq2:SALC} and Figure \ref{fig:sankey} - \textbf{RQ$_{2.1}$}), the requirement-to-architecture (Req-to-Arch) is the most frequently targeted phase, as mentioned in 40\% (24) of the papers. This suggests that LLMs are frequently used to fill in the requirement and architectural design gap,  to assist in mapping textual specifications into formal architectural representations. 
In fact, \wlcite{Ahmad23} leveraged GenAI for collaborative architectural design to assist practitioners in designing the SA from requirements. Similarly, \wlcite{Arias24} used ChatGPT to generate microservice names (architecture) based on the requirements.

Following this, Architecture-to-Code (Arch-to-Code) is also a compelling use case, accounting for 32\% (19) of the research. This indicates a significant focus on using LLMs to automate or help in mapping architectural designs to implementation-level code. Following the same logic, \wlcite{Arun25} used GenAI to generate a serverless function (code) from the architectural specification. However, a peculiar instance and the least explored is Architecture-to-Architecture (Arch-to-Arch) transitions, which only 3\% (2) of the research covers, indicating the lack of current community interest in enhancing, migrating, or converting architectures using LLMs. 	In line with this, \wlcite{Pandini25} refactored the architectural smells using LLMs such as GPT-4 and LLaMA, while~\wlcite{Rubei25} used Gemini 1.5 and GPT-4o to recommend resolutions of architectural violations.

On the other hand, code-to-architecture (8; 13\%) and requirement-to-architecture-to-code (7; 12\%) are fairly represented. 
The former is indicative of efforts toward reverse engineering existing codebases for architectural purposes. Consistent with this approach, \wlcite{Diaz25} experimented with developing LLM-based architecture agents that could improve architecture decision-making starting from code, while \glcite{Fujitsu25} presented its LLM-based tool to perform the architectural reconstruction.

The requirement-to-architecture-to-code illustrates efforts to optimize the entire process from requirements to architecture to code generation. Using this SALC arch, \glcite{Data23} presented in its video tutor an LLM-based copilot of such a SALC arch. Similarly, in a position paper, \wlcite{Raghavan24} presented an assisted architecture LLM based on LLM-based software on LLM-based one requirement.

The distribution of studies indicates that the significant use of LLMs is at the beginning of the SALC, e.g., during requirement analysis as well as architectural design, with less effort going toward changing or reorganizing existing architectures.

\begin{table}
\centering
\footnotesize
\caption{Use of LLMs in the Software Architecture Life Cycle - (RQ$_{2.1}$)}
\label{tab:rq2:SALC}
\resizebox{\linewidth}{!}{%
\begin{tabular}{l p{4cm} rr}
 \hline
\textbf{Code}            & \textbf{PaperID}     & \textbf{Count*} & \textbf{\%} \\ \hline
Req-to-Arch         & \wlcite{Ahmad23, Arias24, Dhar24a, Dhar24b, Diaz24, Eisenreich24, Heiben24, Jahic24, Jose24, Lutze24,   Raghavan24, Rejithkumar24, Supekar24, Tagliaferro25, Tang23, Wei24}, \glcite{Chandraraj23, Data23, Manuel24, Martelli23, Nandi24, Paradkar23, Prakash24, Seroter23} & 24 & 40\%    \\
Arch-to-Code        & \wlcite{Adnan25, Arun25,Duarte25, Fuchb25, Hagel25, Ivers25, Prakash24, Quevedo24, Raghavan24, Saarinen24, Singh24,  Sharma24, Soliman25, Wei24}, \glcite{Ahuja24,  Data23, Martelli23, Nandi24, Paradkar23}                                                               & 19 & 32\%    \\
Code-to-Arch        & \wlcite{Diaz25, Johansson24, Mino24, Ronau24, Rukmono23, Rukmono24, Schindler24}, \glcite{Fujitsu25}                                                                                                                                                                      & 8  & 13\%    \\
Req-to-Arch-to-Code & \wlcite{Raghavan24, Wei24}, \glcite{Data23, Martelli23, Nandi24, Paradkar23, Prakash24}                                                                                                                                                                                   & 7  & 12\%    \\
Arch-to-Arch        & \wlcite{Pandini25, Rubei25}                                                                                                                                                                                                                                      & 2  & 3\%     \\ \hline
\multicolumn{2}{l}{*One paper can have more than one purpose} \\
\end{tabular}%
}
\end{table}

\begin{keyTakeAways}[RQ$_{2.1}$ (SALC Phases)]
LLMs are most frequently applied in the Requirement-to-Architecture (40\%) and Architecture-to-Code (32\%) transitions, while Architecture-to-Architecture (3\%) is the least explored.
\end{keyTakeAways}

\subsubsection{Architectural Styles and \ReviewerB{Patterns} (RQ$_{2.2}$)}
Concerning the architectural styles and patterns to which LLMs have been applied, monolithic architectures are mentioned most frequently, appearing in 15\% (7) of the articles (Table \ref{tab:RQ2:ArchitectureStyle_Patterns}  - \textbf{RQ$_{2.2}$}). This suggests that LLMs are applied primarily in the understanding, analysis, or modernization of monolithic systems. In fact, \wlcite{Rukmono24} used LLM to perform architectural recovery from a legacy monolithic system to understand the program.

As expected, microservices also have a strong appearance and studies investigating their architectural aspects in 6\% (3) of the studies. 

The purpose of preserving the microservice architecture varies. For example, \wlcite{Quevedo24} uses LLM to analyze the code of a microservice-based system to answer architectural questions related to its designs (program comprehension). Similarly, \wlcite{Mino24} focused on the identification of antipatterns in a microservice-based system.

Other trends, such as Self-Adaptive Architecture, Serverless, Layered Architecture, and Model-Based Architecture, only appear erratically, each in 2\% (1) of the studies, showing low research interest in these architectural styles.

An overwhelming 68\% (32) of the research failed to include any data on architectural styles or trends, and it can be inferred that the majority of the work carried out on LLMs within software architecture does not necessarily correlate their conclusions or base the focus on a certain architectural style. 

Such an asymmetrical distribution demonstrates that although the focus is given to some of the architectural schools, especially monolithic and microservices, others are left unexplored regarding the application of LLMs.
\begin{table}
\centering
\caption{Use of LLMs for Architectural Style and Patterns - (RQ$_{2.2}$)}
\label{tab:RQ2:ArchitectureStyle_Patterns}
\resizebox{\linewidth}{!}{%
\begin{tabular}{l p{4.5cm} rr}
 \hline
\textbf{Code}   &\textbf{ PaperID}   & \multicolumn{1}{l}{\textbf{Count*}} & \multicolumn{1}{l}{\textbf{\%}}  \\\hline
Unspecified                       & \wlcite{Ahmad23, Dhar24a, Dhar24b, Diaz24, Diaz25, Eisenreich24, Fuchb25, Heiben24, Ivers25, Johansson24, Jose24, Lutze24,  Pandini25, Prakash24, Raghavan24, Rejithkumar24, Ronau24, Rubei25, Saarinen24, Sharma24, Soliman25, Tagliaferro25, Tang23, Wei24}, \glcite{Ahuja24, Chandraraj23, Data23, Manuel24, Martelli23, Nandi24, Paradkar23, Seroter23} & 32                        & 68\%                      \\
Monolithic                 & \wlcite{Jahic24, Mino24, Rukmono23, Rukmono24, Schindler24, Singh24}, \glcite{Fujitsu25}                                                                                                                                                                                                                                                         & 7                         & 15\%                      \\
Microservices              & \wlcite{Arias24, Duarte25, Quevedo24}                                                                                                                                                                                                                                                                                                             & 3                         & 6\%                       \\
Design Patterns            & \wlcite{Supekar24}                                                                                                                                                                                                                                                                                                                                & 1                         & 2\%                       \\
Layered Architecture       & \wlcite{Rukmono24}                                                                                                                                                                                                                                                                                                                                & 1                         & 2\%                       \\
Model-Based Architecture   & \wlcite{Hagel25}                                                                                                                                                                                                                                                                                                                                  & 1                         & 2\%                       \\
Self-Adaptive Architecture & \wlcite{Adnan25}                                                                                                                                                                                                                                                                                                                                  & 1                         & 2\%                       \\
Serverless                 & \wlcite{Arun25}                                                                                                                                                                                                                                                                                                                       & 1                         & 2 \%   \\                   \hline
\multicolumn{2}{l}{*One paper can have more than one use of LLMs} \\
\end{tabular}%
}
\end{table}

\begin{keyTakeAways}[RQ$_{2.2}$ (Architectural Styles and \ReviewerB{Patterns})]
LLMs mainly target monolithic (15\%) and microservices architectures, with 68\% of studies omitting style details. 
%UML dominates (17\%), while alternatives (2\% each) remain underexplored. Most studies (74\%) lack formal architectural modeling and (87\%) do not contain information on the use of MDE.
\end{keyTakeAways}

\subsubsection{Quality  and Maintenance Tasks (RQ$_{2.3}$)}
Concerning quality aspects,  38\% of the works explicitly discuss antipattern detection using methods such as LLM-based architectural smell refactoring, AI-based detection, and rule-based learning (\textbf{RQ$_{2.3}$}). In particular, \wlcite{Mino24} and~\wlcite{Pandini25} use LLM to detect antipatterns.

Concerning refactoring as a means of removing smells and improving overall software quality, \wlcite{Jose24} and~\wlcite{Supekar24} use LLM to aid in refactoring efforts. Moreover, \wlcite{Ivers25} are the only authors who use an external tool (EM-Assist) to aid in refactoring, in conjunction with LLMs. 

Similarly, studies that perform architectural reconstruction rely on LLM to achieve this. More specifically, \wlcite{Johansson24}  used LLM to map code components to a specific architecture, while~\wlcite{Rukmono24} used LLM to recover the deductive software architecture. Finally, only~\wlcite{Pandini25} reported the use of external tools, validating the observation that LLMs are increasingly being used to recover architectural knowledge and are decreasing in strictly classical tools.

\begin{keyTakeAways}[RQ$_{2.3}$ (Quality \& Maintenance Tasks)]
38\% of studies use LLMs for antipattern detection, refactoring (\wlcite{Jose24}, \wlcite{Supekar24}), and architectural reconstruction (\wlcite{Johansson24}, \wlcite{Rukmono24}). Few integrate external tools, suggesting that LLMs are replacing traditional recovery methods.
\end{keyTakeAways}

\subsubsection{Architecture Modeling and Validation Methods (RQ$_{2.4}$)}
 
Similarly to programming languages, we can represent SA via many architectural languages (AL). Among such AL, UML (Unified Modeling Language) is most commonly applied as a notation in 17\% (8) of the studies (Table \ref{tab:RQ2:ArchitecturalModellingLanguage}   - \textbf{RQ$_{2.2}$}) thus assessing UML as the still dominant modeling language for studies studying LLM due to its versatility in software design and architecture documentation \cite{esposito_large_2024}. For example, \wlcite{Tagliaferro25} used LLM to generate UML component diagrams from informal specifications.

The remaining modeling approaches, i.e., C4, ADR (Architecture Decision Records), SysML, and Knowledge Graphs (KG), each appear only in one study~\wlcite{Dhar24a}  (2\%), indicating little exploration of other architectural modeling notations. In particular, \wlcite{Dhar24a} uses ADR while using LLM to generate architectural design decisions with LLM. In contrast, \wlcite{Heiben24} investigated automating architecture generation using LLMs in Model-Based Systems Engineering using SysML as the modeling language. \glcite{Fujitsu25} used KG for LLM-based architectural reconstruction. Finally, \wlcite{Jahic24} used a combination of UML and C4 for LLM-based assisted architectural decision-making. 

Most of the studies (35; 74\%) did not report any data on the use of architectural modeling languages, suggesting that much research on LLM in software architecture does not necessarily use or elaborate formal modeling approaches. The prevalence of UML and the non-wider deployment of rival model languages suggest there is still sufficient scope for extension research combining LLMs and architected presentation forms.
\begin{table}
\centering
\caption{Architectural Modelling Language - (RQ$_{2.2}$)}
\label{tab:RQ2:ArchitecturalModellingLanguage}
\resizebox{\linewidth}{!}{%
\begin{tabular}{l p{5.5cm} rr}
\hline
\textbf{Code}            & \textbf{PaperID}     & \textbf{Count*} & \textbf{\%} \\ \hline
Unspecified           & \wlcite{Adnan25, Arias24, Arun25, Dhar24b, Diaz24, Diaz25, Duarte25, Eisenreich24, Fuchb25, Hagel25, Ivers25, Johansson24, Jose24, Mino24, Pandini25, Quevedo24, Rejithkumar24, Ronau24, Rubei25, Rukmono23, Rukmono24, Saarinen24, Sharma24, Schindler24, Singh24, Soliman25, Supekar24}, \glcite{Ahuja24, Chandraraj23, Manuel24, Martelli23, Nandi24, Paradkar23, Prakash24, Seroter23} & 35 & 74\%    \\
UML             & \wlcite{Ahmad23, Jahic24, Lutze24, Raghavan24, Tagliaferro25, Tang23, Wei24}, \glcite{Data23} & 8  & 17\%    \\
ADR             & \wlcite{Dhar24a} & 1  & 2\%     \\
C4              & \wlcite{Jahic24} & 1  & 2\%     \\
Knowledge Graph & \glcite{Fujitsu25} & 1  & 2\%     \\
SysML           & \wlcite{Heiben24} & 1  & 2\%     \\ \hline
\multicolumn{2}{l}{*One paper can have more than one language} \\
\end{tabular}%
}
\end{table}

On the topic of architectural design language, five studies reported using some form of Model Driven Engineering (MDE) (Table \ref{tab:RQ2:MDEWho} - \textbf{RQ$_{2.2}$}). More specifically,
\wlcite{Adnan25} used MDE for the generation of the IoT architecture, while~\wlcite{Hagel25} for low-code platform consistency, \wlcite{Supekar24} generated UML component diagrams, \wlcite{Jose24} mapping of the source code to the architecture,\wlcite{Rubei25} provided architectural conformance recommendations, and~\glcite{Rukmono23} deductive architecture recovery, each of which occurs in 2\% (1) of the articles. However, 87\% (40) of the articles did not contain information on the utilization of MDE, so while there is evidence of research that uses LLM for MDE applications, the topic is still fairly unexplored compared to other architectural activities.

\begin{table}
    \centering
    \caption{Model-Driven Engineering (MDE) - (RQ$_{2.2}$)}
    \label{tab:RQ2:MDEWho}
    \resizebox{\linewidth}{!}{%
    \begin{tabular}{p{4.5cm} p{4cm} rr}
        \hline
        \textbf{Code} & \textbf{PaperID} & \textbf{Count} & \textbf{\%} \\
        \hline
        Unspecified & \wlcite{ Ahmad23, Arias24, Arun25, Dhar24a, Dhar24b, Diaz24, Diaz25, Duarte25, Eisenreich24, Fuchb25,  Heiben24, Ivers25, Jahic24, Johansson24, Sharma24,  Lutze24, Mino24, Pandini25, Quevedo24, Raghavan24, Rejithkumar24, Ronau24, Rukmono24, Saarinen24, Schindler24, Singh24, Soliman25, Tagliaferro25, Tang23, Wei24},
        \glcite{Ahuja24,  Chandraraj23, Data23, Fujitsu25, Manuel24, Martelli23, Nandi24, Paradkar23, Prakash24, Seroter23} & 40 & 87\%    \\
        IoT Architecture Generation & \wlcite{Adnan25} & 1  & 2\%    \\
        Low-code Platform\newline Consistency & \wlcite{Hagel25} & 1  & 2\%    \\
        UML Component Diagram Generation & \wlcite{Supekar24} & 1  & 2\%    \\
        Source Code to Architecture Mapping & \wlcite{Jose24} & 1  & 2\%    \\
        Architectural Conformance Recommender & \wlcite{Rubei25} & 1  & 2\%    \\
        Deductive Software \newline Architecture Recovery   & \wlcite{Rukmono23} & 1  & 2\%    \\
        \hline
    \end{tabular}%
    }
\end{table}
93\% (43) of the studies report that no information was provided on the LLM model output validation techniques (Table \ref{tab:RQ2:ArchAnalysisValidation} - \textbf{RQ$_{2.4}$}) while only three of them report how they evaluated the LLM model output. In particular, ~\wlcite{Eisenreich24} used ATAM (Architecture Tradeoff Analysis Method), while~\wlcite{Ahmad23} used SAAM (Software Architecture Analysis Method) and~\wlcite{Quevedo24} used static analysis. Hence, our findings suggest that formal assessment methods are still not in common practice, and most studies do not explicitly validate their AI-generated architectural designs. 
\begin{table}
    \centering
    \footnotesize
    \caption{Architecture Analysis Method - Adopted Generative AI Outputs Validation Methods - (RQ$_{2.3}$)}
    \label{tab:RQ2:ArchAnalysisValidation}
    \resizebox{\linewidth}{!}{
    \begin{tabular}{l p{4cm} rr}
        \hline
        \textbf{Code} & \textbf{PaperID} & \textbf{Count} & \textbf{\%} \\
        \hline
        Unspecified & \wlcite{Adnan25, Arias24, Arun25, Dhar24a, Dhar24b, Diaz24, Diaz25, Duarte25, Fuchb25, Hagel25, Heiben24, Ivers25, Jahic24, Johansson24, Jose24, Lutze24, Mino24, Pandini25, Raghavan24, Rejithkumar24, Ronau24, Rubei25, Rukmono23, Rukmono24, Saarinen24, Sharma24, Schindler24, Singh24, Soliman25, Supekar24, Tagliaferro25, Tang23, Wei24}, \glcite{Ahuja24, Chandraraj23, Data23, Fujitsu25, Manuel24, Martelli23, Nandi24, Paradkar23, Prakash24, Seroter23}  & 43 & 93\%    \\
        ATAM & \wlcite{Eisenreich24} & 1  & 2\%    \\
        SAAM & \wlcite{Ahmad23} & 1  & 2\%    \\
        Static Analysis & \wlcite{Quevedo24} & 1  & 2\%    \\
        \hline
    \end{tabular}%
    }
\end{table}

\begin{keyTakeAways}[RQ$_{2.4}$ (Modeling \& Validation Methods)]
The most used architecture modeling language is UML (17\%), while alternatives (2\% each) remain underexplored. Most studies (74\%) lack formal architectural modeling and (87\%) do not contain information on the use of MDE.
ATAM, SAAM, and static analysis are the only validation methods reported, while 93\% of the studies do not report any evaluation strategy, indicating a lack of systematic validation for AI-generated architectural output.
\end{keyTakeAways}

\subsubsection{Generative AI for Software Architecture: In which cases (RQ$_{2.5}$)}
This subsection presents the specific use cases in which GeneAI has been applied to the software architecture. We examine the types of systems analyzed, the domains in which LLMs are deployed, and the programming languages associated with these use cases. 
Table \ref{tab:RQ2:UseCases} presents the use cases and systems addressed in the research papers that apply GenAI to software architecture. According to Table \ref{tab:RQ2:UseCases}, Requirements and Architectural Snippets are the most common subjects, appearing in 15\% (7) of research papers, which indicates that LLMs are widely tested in fragments of architectural information \wlcite{Dhar24b, Rejithkumar24}. Enterprise and Property Software and IoT, and Smart Systems also attract significant interest, indicating applications in industrial and network environments. For example, \wlcite{Singh24} used LLMs to reengineer a legacy system at Volvo Group.
Since it is challenging to retrieve large-scale open source systems or to evaluate prioritized mobile applications and embedded systems, our findings evidenced that such domains are underrepresented in our study. For example, \glcite{Ahuja24} experimented with RAG to evaluate green software patterns starting from architectural documents from Instagram, WhatsApp, Dropbox, Uber, and Netflix. Similarly, \wlcite{Rukmono24} investigated the architectural reconstruction of an Android app.
Finally, 38\% (18) of the research articles did not specify a precise use case, that is, position or vision articles.

\begin{table*}[t]
    \centering
    \caption{Use Cases and Systems Analyzed - (RQ$_{2.5}$)}
    \label{tab:RQ2:UseCases}
    \resizebox{\linewidth}{!}{%
    \begin{tabular}{l p{4.5cm} rr}
        \hline
        \textbf{Category} & \textbf{PaperID} & \textbf{Count*} & \textbf{\%} \\
        \hline
        \textbf{Unspecified} & \wlcite{Diaz24, Diaz25, Eisenreich24, Ivers25, Jahic24, Mino24, Raghavan24, Ronau24, Sharma24, Supekar24}, \glcite{Chandraraj23, Data23, Fujitsu25, Manuel24, Martelli23, Nandi24, Paradkar23, Prakash24} & 18 & 38\%   \\
        \textbf{Requirement and Architectural Snippets} & \wlcite{Arias24, Arun25, Dhar24a, Dhar24b, Rejithkumar24, Schindler24, Rukmono23}& 7 & 15\%   \\
        \textbf{Social Media and Large-Scale Systems} & \glcite{Ahuja24} & 1 & 2\%   \\
        \quad Architectural documents of Instagram, WhatsApp, Dropbox, Uber, Netflix &  &  \\
        \textbf{Educational and Research Platforms} & \wlcite{Fuchb25} & 1 & 2\%   \\
        \quad BigBlueButton, JabRef, TEAMMATES, TeaStore &  &  \\
        \textbf{Cloud and Open-Source Solutions} & \wlcite{Soliman25, Fuchb25, Pandini25}, \glcite{Seroter23} & 4 & 8\%  \\
        \quad Google Jump-Start Solution, Hadoop HDFS, MediaStore, Multiple Open-Source Projects &  &  \\
        \textbf{IoT and Smart Systems} & \wlcite{Rubei25, Adnan25, Lutze24, Heiben24} & 4 & 8\% \\
        \quad IoT Reference Architectures, Smart City IoT System, Smartwatch App, Remote-Controlled Autonomous Car &  &  \\
        \textbf{Mobile and Layered Applications} & \wlcite{Rukmono24} & 1 & 2\%   \\
        \quad Layered App (Android) &  &  \\
        \textbf{Low-Code and Microservices Architectures} & \wlcite{Hagel25, Duarte25, Quevedo24} & 3 & 6\%   \\
        \quad Low-Code Development Platforms, Microservices in GitHub, TrainTicket Microservice Benchmark &  &  \\
        \textbf{Monolithic and Traditional Architectures} & \wlcite{Ahmad23} & 1 & 2\%   \\
        \quad Monolithic, Single Component &  &  \\
        \textbf{Enterprise and Proprietary Software} & \wlcite{Jose24, Saarinen24, Wei24, Singh24}& 4 & 8\%   \\
        \quad Proprietary Enterprise Scenarios, Ordering System, SuperFrog Scheduler, Volvo SCORE System &  &  \\
        \quad Requirement Snippets, Snippets of Code, Snippet of Architectural Design Records, Architectural Snippets &  &  \\
        \textbf{Automotive and Embedded Systems} & \wlcite{Johansson24} & 1 & 2\%   \\
        \quad PX4 (Drone Software) &  &  \\
        \textbf{Text-Based and Specialized Systems} & \wlcite{Tang23, Tagliaferro25} & 2 & 4\%   \\
        \quad Text/Aviation System, Software Engineering Exam Traces &  &  \\
        \hline
    \multicolumn{2}{l}{*One paper can have more than one use case} \\
    \end{tabular}%
    }
\end{table*}
% \begin{keyTakeAways}[RQ$_{2.5}$ (Use Cases)]
% LLMs are most frequently applied to architectural requirements and snippets (15\%), with notable usage in enterprise software and IoT systems (8\%), while large-scale, mobile, and embedded systems are less explored.
% \end{keyTakeAways}

Table \ref{tab:RQ3:UCPL} presents the programming languages of the use cases examined. As is evident from Table \ref{tab:RQ3:UCPL}, the most frequent language is Java (7; 13\%), reflecting that Java systems are leading the research on LLM applications in software architecture. Other languages, including JavaScript, Python, UML, and Natural Language (NL), occur to a smaller extent, reflecting a mix of implementation and design-level notation.

A significant 58\% (30) of the articles did not report the programming language of the use case, and this is an area of reporting that hinders the measurement of LLM uptake by the technology stacks. The presence of legacy languages such as COBOL (1; 2\%) suggests that there is research on legacy systems, but only in a very limited subset of cases. These results show that although Java is the most mentioned language, there is no domination of any language, and the granularity of implementation decision details differs among studies.
\begin{table}
\centering
\caption{Use Case Programming Language - (RQ$_{2.5}$)}
\label{tab:RQ3:UCPL}
\resizebox{\linewidth}{!}{%
\begin{tabular}{l p{6.2cm} rr}
\hline
\textbf{Code}     & \textbf{PaperID}     & \textbf{Count} & \textbf{\%} \\ \hline
Unspecified        & \wlcite{Adnan25, Diaz24, Diaz25, Duarte25, Eisenreich24, Fuchb25, Hagel25, Heiben24, Ivers25, Jahic24, Jose24, Lutze24, Mino24, Prakash24, Raghavan24, Ronau24, Rubei25, Sharma24,Singh24, Soliman25, Supekar24, Tagliaferro25}, \glcite{Ahuja24, Chandraraj23, Data23, Manuel24, Martelli23, Nandi24, Paradkar23, Seroter23} & 30 & 58\% \\
Java       & \wlcite{Pandini25, Quevedo24, Rukmono23, Rukmono24, Saarinen24, Schindler24}, \wlcite{Arun25} & 7  & 13\%  \\
Nature Language         & \wlcite{Arias24, Dhar24a, Dhar24b, Rejithkumar24} & 4  & 8\%  \\
JavaScript & \wlcite{Saarinen24}, \wlcite{Arun25} & 2  & 4\%  \\
Python     & \wlcite{Wei24}, \wlcite{Arun25} & 2  & 4\%  \\
UML        & \wlcite{Ahmad23, Tang23} & 2  & 4\%  \\
C++        & \wlcite{Johansson24} & 1  & 2\%  \\
COBOL      & \glcite{Fujitsu25} & 1  & 2\%  \\
Node.js    & \wlcite{Saarinen24} & 1  & 2\%  \\
React      & \wlcite{Saarinen24} & 1  & 2\%  \\
TypeScript & \wlcite{Arun25} & 1  & 2\%  \\ \hline
\multicolumn{4}{l}{*One paper can have more than one Programming Language} \\
\end{tabular}%
}
\end{table}
\begin{keyTakeAways}[RQ$_{2.5}$ (Use Cases)]
LLMs are most frequently applied to architectural requirements and snippets (15\%), with notable usage in enterprise software and IoT systems (8\%), while large-scale, mobile, and embedded systems are less explored.
Moreover, Java (13\%) is the language most commonly used in LLM-driven architectural studies, but 58\% of the studies do not specify a programming language, highlighting a gap in reporting on implementation details.
\end{keyTakeAways}

\subsection{Generative AI for Software Architecture: Future Challenges (RQ$_{3}$)}
\label{sec:RQ3}
This subsection presents key challenges identified in the original studies, highlighting limitations in model reliability, ethical concerns, quality of AI-generated outputs, and practical integration issues, which need to be addressed for broader adoption.

Future challenges in GenAI research for SA primarily include the \textbf{accuracy} of LLM (9; 15\%), the most cited issue, emphasizing the difficulty in maintaining accurate and reliable model outputs. LLM \textbf{hallucinations} (5; 8\%) also represent a critical issue, necessitating mechanisms to prevent incorrect or misleading responses (Table \ref{tab:RQ4:FutureChallenges} - \textbf{RQ$_4$}).

\textbf{Ethics}-related concerns (4; 7\%), \textbf{privacy/data privacy}  (7; 12\%), and \textbf{human interaction} with LLM (3; 5\%) indicate that aligning AI outputs with responsible and interpretable practices is increasingly important. Specifically, \glcite{Manuel24} highlights ethical considerations as a major challenge, with issues such as bias in AI-generated architectural decisions and lack of transparency in model reasoning posing significant concerns, particularly in critical domains like healthcare or finance. \glcite{Nandi24} and \glcite{Martelli23} additionally emphasize privacy risks associated with inadvertent information leakage, advocating stronger data protection mechanisms. Addressing these ethical and privacy challenges requires regulatory frameworks, improved model interpretability, and robust security measures.

Moreover, the absence of standardized and task-specific evaluation metrics remains a significant barrier, hindering systematic validation and refinement of AI-generated architectural decisions. Limited context-awareness due to fragmented and inconsistent inputs and inadequate explainability mechanisms further compounds the challenges. Current GenAI models struggle with long-context reasoning and inherent uncertainties such as rigidity in post-hoc adjustments of outputs, persistent hallucinations, and bias, which compromise reliability and trust.

Architectural degradation risks due to overuse or blind trust in AI-generated recommendations necessitate rigorous human oversight and verification processes. Additionally, there is a clear need for architecture-specific datasets, benchmarks, and clearer \textbf{semantic traceability} between architectural artifacts to rigorously test and compare model effectiveness.

\textbf{Quality of generated code}, \textbf{maintainability} (2; 4\%) \wlcite{Sharma24,Adnan25}, \textbf{scalability}, and \textbf{security} (2; 4\%) \wlcite{Arun25,Chandraraj23} represent important areas needing attention. LLM output generalizability (2; 4\%) \wlcite{Ronau24,Arias24}, \textbf{reduced human creativity} (2; 4\%) \glcite{Paradkar23,Chandraraj23}, \textbf{pattern recognition accuracy} (2; 4\%) \wlcite{Schindler24}, and \textbf{intellectual property} concerns (1; 2\%) \wlcite{Ahmad23} also pose significant challenges that must be explicitly considered.

\textbf{Formal verification} and compliance checking have also been suggested as necessary steps to ensure that AI-generated outputs meet defined architectural and regulatory standards \glcite{Paradkar23}.
In general, studies highlight accuracy, hallucinations, ethics, and practical integration as critical concerns, suggesting a strong need for systematic validation approaches for the generated architectural solutions, enhanced explainability, and rigorous benchmarks for future GenAI adoption in software architecture.
\begin{table}
\centering
\caption{Future Challenges - (RQ$_3$)}
\label{tab:RQ4:FutureChallenges}
\resizebox{\linewidth}{!}{%
\begin{tabular}{l p{4cm} rr}
\hline
\textbf{Code}            & \textbf{PaperID}     & \textbf{Count*} & \textbf{\%} \\ \hline
LLM Accuracy                         & \wlcite{Arun25,Fuchb25, Johansson24, Jose24, Lutze24, Soliman25}, \glcite{ Nandi24, Manuel24, Martelli23}  & 9  & 15\% \\
Unspeficied                                  & \wlcite{Singh24, Rejithkumar24, Dhar24b, Rukmono23, Mino24, Jahic24}, \glcite{Data23, Seroter23, Fujitsu25} & 9  & 15\% \\
LLM Hallucinations                   & \wlcite{Diaz24, Saarinen24, Tang23}, \glcite{Ahuja24, Manuel24}                                            & 5  & 8\%  \\
Ethical Considerations               &  \glcite{Nandi24, Manuel24, Martelli23, Ahmad23}                                                   & 4  & 7\%  \\
Privacy                              & \wlcite{Tang23}, \glcite{Nandi24, Manuel24, Martelli23}                                                    & 4  & 7\%  \\
Architectural Solution Validation    & \wlcite{Eisenreich24, Rukmono24, Johansson24, Schindler24}                                                     & 4  & 7\%  \\
Data Privacy                         & \wlcite{Tang23}, \glcite{Nandi24, Martelli23}                                                              & 3  & 5\%  \\
Generated Code Maintenability        & \wlcite{Quevedo24, Sharma24}                                                              & 2  & 4\%  \\
Generated Code Quality               & \wlcite{Raghavan24, Dhar24a, Johansson24}                                                         & 3  & 5\%  \\
LLM Human Interaction                & \wlcite{Eisenreich24, Supekar24, Ahmad23}                                                         & 3  & 5\%  \\
Traceability                         & \wlcite{Heiben24,  Wei24}    \glcite{Prakash24}                                                           & 3  & 5\%  \\
Generated Code Security              & \wlcite{Arun25} \glcite{Chandraraj23}                                                                     & 2  & 4\%  \\
LLM Output Generalizability          & \wlcite{Ronau24, Arias24}                                                                         & 2  & 4\%  \\
Reduced Human Creativity             & \glcite{Chandraraj23, Paradkar23}                                                                 & 2  & 4\%  \\
Pattern Recognition Accuracy         &  \glcite{Ahuja24}                                                                                  & 1  & 2\%  \\
Intellectual Property                & \wlcite{Ahmad23}                                                                                  & 1  & 2\%  \\
\hline
\multicolumn{4}{l}{*One paper can have more than one future challenge} \\
\end{tabular}%
}
\end{table}
Surprisingly, 15\% (9) of studies did not mention future challenges explicitly, indicating gaps in recognizing and addressing limitations of LLMs in software architecture.

\begin{keyTakeAways}[RQ$_{3}$ (Future Challenges)]
LLM accuracy (16\%), hallucinations (9\%), and ethical concerns (7\%) dominate, alongside critical challenges in evaluation metrics, context-awareness, explainability, systematic validation methods, generalizability, intellectual property, and formal verification.
\end{keyTakeAways}

\section{Discussion}
\label{sec:Discussion}
This section discusses the challenges implied by or highlighted in the identified literature and elaborates on future directions. Additionally, it summarizes the different implications identified in white and gray literature. Moreover, to complement the discussion of challenges and implications, we analyzed how these aspects are related. 

\subsection{Implications}
\label{sec:Implications}

\ReviewerA{The assessed literature has various implications and suggests future direction as follows:}

\textbf{AI-assisted programming:} It is an excellent opportunity for short-term future direction. Yet, the products have to be explainable, especially in terms of architecture decisions; this correlates with the need for AI products to generate models or graphs like UML sketches to explain to practitioners the proposed products. There are multiple directions and implications we will elaborate on \wlcite{Ronau24}.

\ReviewerA{\textbf{Integration across SALC phases:} Advancement can be claimed once a complete, single integrated GenAI for engineering product development engages in all SALC phases. Currently, we see pieces of the puzzle not necessarily related to the previous phases. An advancement would be to create a framework guiding the integration of all the various tools contributing to one entity or process \glcite{Prakash24}.}

\ReviewerA{\textbf{Evolution, Continuous Architecture, Integration with DevOps:}
Once we deploy GenAI to manage code, there must be reinforcement learning for architecture optimization, and this must take into account the current trends in software systems, such as cloud-native that employs decentralized architecture \cite{Lelovic2024}. Future perspectives might consider tooling that adjusts the systems to their usage, integrating with DevOps by monitoring user requests and trends by tracing and taking into account available hardware resources or their financial costs. However, GenAI support for system evolution must cope with hallucinations, architecture degradation, and given that we are currently dealing with the pieces of a puzzle with GenAI tools rather than a comprehensive framework for the complete SALC, there is a long path to this.}

\ReviewerA{\textbf{Legacy Systems:} Some major challenges for GenAI are the inability to pivot legacy applications and systems \glcite{Martelli23, Nandi24}. Training of GenAI on data and training input is threatened by privacy and intellectual property violations \glcite{Martelli23, Nandi24}.}

\ReviewerA{\textbf{Documentation might become legacy:} While writing documentation can be expedited by GenAI \wlcite{Saarinen24}, will this still be needed in the future? AI can provide interactive documentation by reverse engineering the code or using other static analysis approaches like those presented by Quevedo et al. \wlcite{Quevedo24}. Currently, documentation generation requires human intervention to ensure the usefulness, correctness, and validity of the text, and hallucination in evolving systems can be difficult to overcome \wlcite{Quevedo24}.}

\textbf{Cross-team Decentralized Collaboration with AI:} Future vision must be elaborated on human-centered cross-team collaboration. For instance, in microservices, we deal with a lot of co-changes that involve various teams \cite{Lelovic2024}. One cannot ignore the fact that most current systems run on a decentralized architecture connecting codebases, where consistency is essential when changes take place to limit ripple effects.
Moreover, many issues emerging from the GenAI impact are caused by the neglect of the socio-technical problems and human needs and values \wlcite{Saarinen24}. 
Could GenAI facilitate communication across teams when co-changes must take place? Chandraraj \glcite{Chandraraj23} suggests that GenAI might overlook team dynamics and organizational culture in its architectural suggestions. For example, it might propose a complex solution without considering the team's abilities or the availability of developers with technical skills. It could also suggest a solution that technically works but doesn't align with the organization’s broader objectives.

\textbf{Who manages the generated content:} Model-driven development had one core problem: no one wanted to manage the code that was generated, and when one did, the model generation would not work when the system evolves, as it would override the changes. Similar questions should be asked regarding AI-generated code \glcite{Prakash24},\wlcite{Wei24}. Experimental productivity and quality comparison studies between human-generated and AI-generated code in a
realistic environment are needed \wlcite{Saarinen24}. We need to prevent architectural degradation, and thus, architectural metrics need to be in place.

% \ReviewerA{There are few \textbf{overlaps between the gray and white literature} when it comes to implications and future work:}

\ReviewerA{\textbf{Complex Architectures:}
The progress in GenAI should move towards complex architecture. While facing limitations with tokens, advancing RAG, multi-agent frameworks are inevitable to support and complex architectural styles such as microservices \wlcite{Arun25}.}

\ReviewerA{\textbf{Heterogeneous Sources of input: } such as software architecture documents, tabular data, and technical diagrams, must be considered since most works focus on a single source of truth, and we might more and more consider aspect oriented programming to come into play with GenAI ~\glcite{Ahuja24},\cite{Cerny21042019}.}

\textbf{Formal Verification and AI-Driven Compliance Checking:} 
It is easy to start using GenAI tools; however, maximizing the potential can be challenging \wlcite{Saarinen24}. Moreover, it might be difficult to control the tools, and results need to be checked, as practitioners can easily accept suggestions relying on AI as an oracle. Still, there were observed limits of GenAI to complex tasks with resulting products in less usable \wlcite{Saarinen24}. This leads to comprehension issues, which we mentioned with challenges to generate UML-like models or diagrams to guide developers on explainability. 
\ReviewerA{The correctness, completeness, and effectiveness of the generated code are open to improvement \wlcite{Sharma24},\glcite{Martelli23}. Similar to the mentions in white literature, formal verification or result validation is a worthy direction of research.}

\textbf{Replacement of Human Experts:} 
Literature often mentions that the discipline will move towards a field where human experts manage projects, where GenAI agents can prototype or deliver tasks for them to manage \wlcite{Saarinen24}. This suggests the opportunity for research on AI tools for project management.
AI replacing humans can be approached once we overcome trust and establish evaluation metrics. For instance, Prakash \glcite{Prakash24} suggests GenAI helps developers by that 25\% to write code efficiently, fix bugs, and improve software quality. 
GenAI, as a tool for architects, is not a replacement \glcite{Chandraraj23,Data23}, as it is not yet ready, lacking in complex business contexts, and failing in subjective judgment, business intuition, and personal accountability.
\ReviewerA{Apart from this, Fujitsu \glcite{Fujitsu25} recognizes the need for interactive capabilities to verify current application specifications and assess the impact of source code changes. However, it is important to be aware of the challenges and ethical considerations associated with GenAI \glcite{Paradkar23}. AI algorithms are trained on data, and this data can be biased. This bias can be reflected in the output of AI models. GenAI tools can make mistakes, and they should not be used to replace human judgment. It is also essential to consider the ethical implications of AI-generated architectural patterns and designs before using them.}

% \ReviewerA{However, there are also separable topics for implications for future work to be found in \textbf{gray literature}:}

% \begin{ReviewerCEnv}
% \begin{figure*}[t]
% \centering
% \includegraphics[width=0.9\linewidth]{Figures/Tax.pdf}
% \caption{Graphical representation of the relationships between identified challenges and proposed future directions.}
% \label{fig:challenges-directions-diagram}
% \end{figure*}

% % \todo[inline] {Figure 5: le frecce che escono da Architectural degradation sono troppo simili come colore a quelle che escono d finetuning. 

% % swappa il colore con quello sopra: 
% % Architectural Degradation in Verde
% % Context awareness in azzurro. }
% % \todo[inline]{Typo nella figura: Foamrl method, genrated reuslt, ... ce ne sono tantissime. 
% % Capitalize words!}

% To complement the discussion of challenges (RQ$_3$, Section~\ref{sec:RQ3}) and implications (Section~\ref{sec:Implications}), we analyzed how these aspects are related. This section presents and motivates these relationships based on the evidence from our results (Section~\ref{sec:results}).

\subsection{Relationships Among Challenges}

The challenges identified in Generative AI (GenAI) for Software Architecture (SA) are interconnected, influencing each other and impacting the effectiveness and adoption of these technologies. At the core, \textbf{LLM accuracy} and \textbf{hallucinations} represent foundational issues, significantly affecting the reliability of GenAI outputs. Without addressing these critical concerns, the usefulness of GenAI in SA is limited, as inaccurate or misleading outputs directly undermine trust and adoption.

\textbf{Ethical considerations}, including potential biases and transparency in decision-making processes, closely interact with \textbf{privacy} and \textbf{data privacy} challenges, highlighting a shared necessity for secure and responsible use of AI models. Ethical issues and privacy concerns demand a holistic approach, incorporating rigorous standards, compliance, and transparency mechanisms.

Similarly, the \textbf{validation of architectural solutions} and \textbf{traceability} concerns underline the importance of systematic evaluation metrics and clear semantic relationships among architectural artifacts. Effective \textbf{validation} and \textbf{traceability} practices directly contribute to improved \textbf{maintainability}, \textbf{security}, and \textbf{quality of generated code}. These aspects are integral to maintaining the long-term health and scalability of software systems developed using GenAI.

The \textbf{generalizability} of LLM outputs also intersects significantly with accuracy, as enhanced generalizability demands higher context-awareness and improved pattern recognition capabilities. However, excessive reliance on automated solutions introduces the risk of \textbf{reduced human creativity}, emphasizing the need for balanced human interaction and oversight.

\textbf{Intellectual property} concerns further underline the necessity of clear guidelines and verification methods to ensure that generated outputs comply with existing legal frameworks, preventing misuse and infringement.

In summary, these challenges collectively highlight the necessity for integrated solutions that comprehensively address accuracy, ethics, privacy, validation, and human interaction, fostering reliable, secure, and ethically responsible application of GenAI technologies in software architecture.

\subsection{Relationships Between Challenges and Future Directions}
\label{sec:Future}

 \begin{ReviewerCEnv}

\begin{figure*}[ht]
\centering
\includegraphics[width=\linewidth]{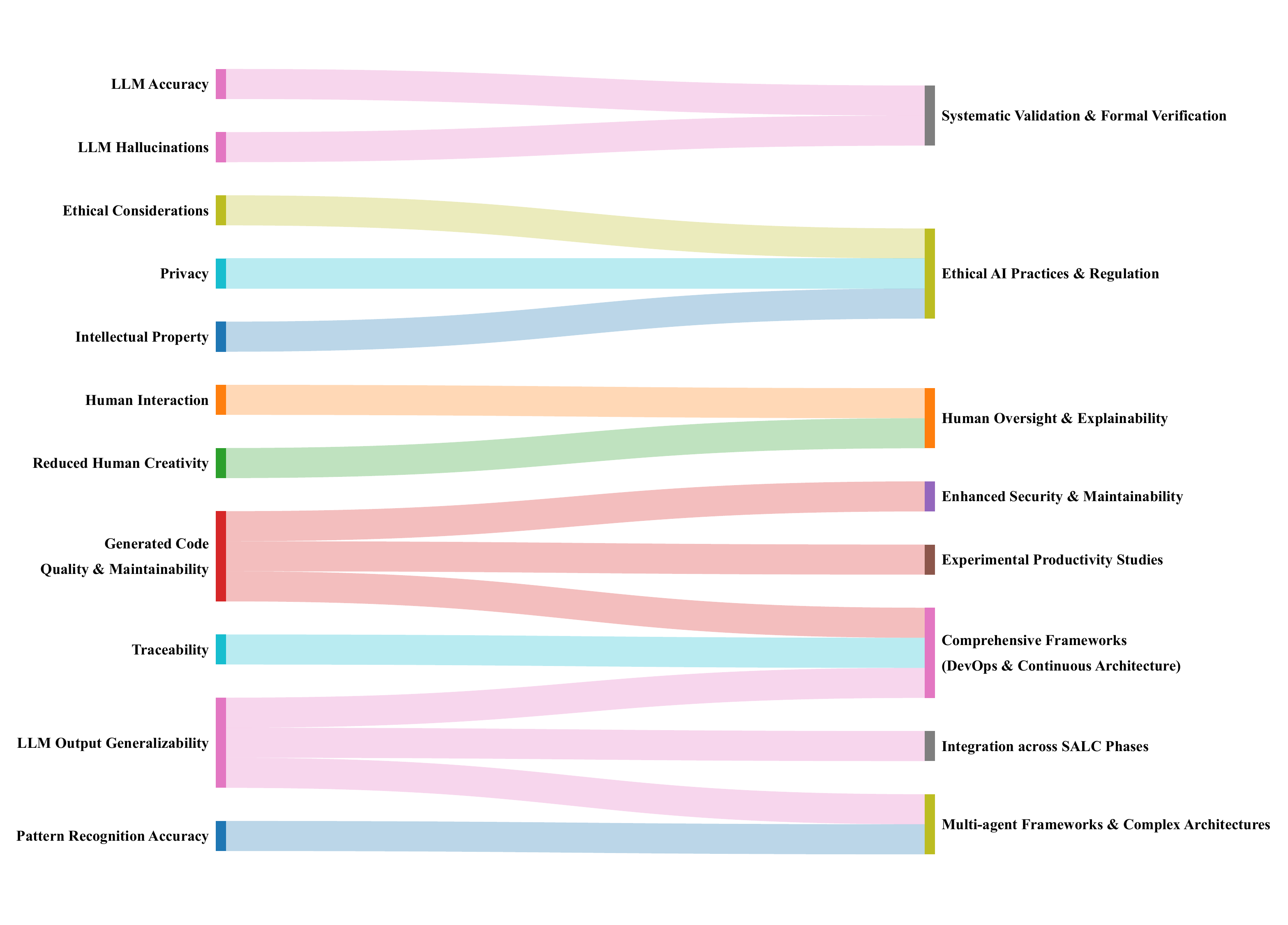}
\caption{Challenges and Implications in GenAI for Software Architecture}
\label{fig:challenges_implications}
\end{figure*}

% visually illustrates the relationships between the identified challenges and their implications.

The identified challenges from the MLR highlight key concerns influencing the future directions of GenAI within software architecture. Addressing these challenges can guide the practical implications for software architecture practice and research (Figure~\ref{fig:challenges_implications}).

The primary challenges emphasized in the literature involve accuracy and reliability, specifically the frequent issue of LLM accuracy and hallucinations. The implications of these challenges point directly to the necessity of enhancing systematic validation approaches for AI-generated architectural solutions, including architectural solution validation. The requirement for precise and robust validation frameworks emerges clearly as LLMs' outputs significantly influence architectural decisions. Formal verification and AI-driven compliance checking have been suggested explicitly to mitigate risks arising from incorrect or misleading AI recommendations.

Ethical considerations and privacy are also prominently discussed challenges, highlighting the importance of integrating ethical AI practices. Concerns about biases, transparency, and inadvertent data leakages underscore the critical need for regulatory frameworks, improved model interpretability, and stronger data protection mechanisms. Ethical and privacy implications strongly suggest the necessity of considering human values and socio-technical dynamics when integrating GenAI tools into architectural practices.

Challenges related to human interaction with LLM, reduced human creativity, and intellectual property concerns further highlight the delicate balance required between automation and human oversight. GenAI must support rather than replace human expertise, as indicated by implications calling for AI-assisted programming to generate explainable artifacts like UML diagrams. This aligns with practical needs, emphasizing that architectural documentation generation currently requires human intervention to ensure accuracy and usability.

The challenge of maintaining the quality of generated code, maintainability, scalability, and security stresses the critical role of human verification processes in the short-term future, particularly when GenAI supports tasks traditionally managed by humans. The implication here is a clear call for experimental productivity and quality studies comparing human-generated and AI-generated code in realistic environments to better understand trade-offs and risks, particularly regarding architectural degradation. Generated code quality and maintainability have implications for comprehensive frameworks integrating DevOps and continuous architecture.

Furthermore, the highlighted challenges related to the generalizability of LLM outputs and traceability suggest the need for enhanced context-awareness and semantic traceability. Addressing these issues aligns with the implications around integrating GenAI across the SALC phases, requiring comprehensive frameworks that manage complex architectures, including continuous architecture and DevOps integration. Specifically, the generalizability of LLM outputs has direct implications for multi-agent frameworks and complex architectures.

Lastly, the challenge of pattern recognition accuracy and managing heterogeneous sources of inputs (e.g., handling multimodal data such as UML diagrams, text in natural language, source code) aligns with implications emphasizing the need for multi-agent frameworks and RAG to handle complex, decentralized architectures such as microservices effectively.

In summary, addressing these highlighted challenges necessitates systematic validation methods, regulatory frameworks, ethical considerations, enhanced human oversight, rigorous comparative studies, and comprehensive integration frameworks. Integrating these insights will be pivotal for effectively realizing the potential of GenAI in software architecture.

\end{ReviewerCEnv}

\begin{ReviewerAEnv}
\subsection{Differences between white and gray literature findings}

Our study, being an MLR, covered both the white and gray literature to explore GenAI for Software Architecture. The findings revealed notable differences between these two sources. More specifically, the white literature, including peer-reviewed conference papers and journal papers, mainly addresses formalizing and generalizing the contribution of LLMs to formal software architecture processes. The white literature focused on \textbf{LLMs to automate or facilitate architectural decision making}, \textbf{traceability}, and model-driven development. Such studies tend to present systematic experiments, propose new methods, or present conceptual foundations to bring LLM into software architecture activities. Moreover, it tends to investigate empirical aspects of LLM use, such as how good they are at generating architectural fragments or determining architectural conformance to predefined standards.

The gray literature comprises blog posts, industry reports, preprints, and white papers and has a more pragmatic and timely focus. LLMs are typically being researched as \textbf{work productivity tools} in contrast to scientific objects of intense investigation. Many sources in the gray literature portray LLMs as assistants that assist in making ongoing software development efforts more straightforward, that is, \textbf{ architecture reconstruction, mapping requirements to architectures, and generating documentation}. The ability of LLMs to act as architectural design copilots, providing quick recommendations or insight versus delving deeper into analytical reasoning, is predominantly what these resources highlight. In contrast to white literature, gray literature features industry-led use cases, for example, using LLMs to plan modernization, automate software lifecycles, and extract knowledge from current codebases.

As expected, the main difference is in the \textbf{assessment approach}: the white literature rigorously analyzes the performance of LLM through empirical research, controlled experiments, and case studies, while the gray literature must suffice with anecdotal evidence or high-level summaries without formal endorsement. Moreover, the white literature is more interested in probing theoretical questions, such as the interpretability and trustworthiness of architectural knowledge generated by LLM. In contrast, gray literature tends to be positive and introduces LLMs as enablers without critically addressing their limitations.

In general, both types of literature promote knowledge of LLM implementation in software architecture but differ concerning the purpose and level of critique. The white literature is more research-focused and methodologically clear, and its purpose is to refine and establish LLM integration within the architecture process. The gray literature offers a rapid path to industry learning, whose goal is adoption, tool reviews, and short-term benefit. 
Since technological hype is a mixture of academic and industry interests, we performed this MLR to capture both worlds and to present a complementary view on the state of the art.

\end{ReviewerAEnv}
\section{Threats to Validity}
\label{sec:threats}
The results of an MLR may be subject to validity threats, mainly concerning the correctness and completeness of the survey.
We have structured this Section as proposed by Wohlin et al.~\cite{Wohlin2014}, including construct, internal, external, and conclusion validity threats.

\textbf{Construct validity}.
Construct validity is related to the generalization of the result to the concept or theory behind the study execution~\cite{Wohlin2014}. In our case, it is related to the potentially subjective analysis of the selected studies.
As recommended by Kitchenham’s guidelines~\cite{Kitchenham2007}, data extraction was performed independently by two or more researchers and, in case of discrepancies, a third author was involved in the discussion to clear up any disagreement. Moreover, the quality of each selected paper was checked according to the protocol proposed by Dyb{\aa} and Dings{\o}yr~\cite{Dyba2008}. 

\textbf{Internal validity}.
Internal validity threats are related to possible wrong conclusions about causal relationships between treatment and outcome~\cite{Wohlin2014}. In the case of secondary studies, internal validity represents how well the findings represent the findings reported in the literature. To address these threats, we carefully followed the tactics proposed by~\cite{Kitchenham2007}.

\textbf{External validity}.
External validity threats are related to the ability to generalize the result~\cite{Wohlin2014}. In secondary studies, external validity depends on the validity of the selected studies. If the selected studies are not externally valid, the synthesis of its content will not be valid either. In our work, we were not able to evaluate the external validity of all the included studies.

\textbf{Conclusion validity}.
Conclusion validity is related to the reliability of the conclusions drawn from the results~\cite{Wohlin2014}. In our case, threats are related to the potential non-inclusion of some studies. To mitigate this threat, we carefully applied the search strategy, performing the search in eight digital libraries in conjunction with the snowballing process~\cite{Wohlin2014}, considering all the references presented in the retrieved papers, and evaluating all the papers that reference the retrieved ones, which resulted in one additional relevant paper.  We applied a broad search string, which led to a large set of articles, but enabled us to include more possible results. We defined inclusion and exclusion criteria and applied them first to the title and abstract. However, we did not rely exclusively on titles and abstracts to establish whether the work reported evidence of architectural degradation. Before accepting a paper based on title and abstract, we browsed the full text, again applying our inclusion and exclusion criteria.

\section{Conclusions}
\label{sec:conclusions}
This study presents the results of a multivocal review of the literature investigating the topic of LLM and GenAI applications in the domain of software architecture. It investigated the various perspectives of such practices, including the rationales for applying different LLM models and approaches, application contexts in the software architecture domain, use cases, and potential future challenges. From four well-recognized academic literature sources and the three most popular search engines, it extracted 36 white literature and 10 gray literature. 

The analyzed results show that LLMs have mainly been applied to support architectural decision-making and reverse engineering, with the GPT model being the most widely adopted. Meanwhile, few-shot prompting is the most commonly adopted technique when human interaction is involved in most studies. Requirement-to-code and Architecture-to-code are the SALC phases where LLMs are mostly applied, while monolith and microservice architectures are the ones that draw the most attention in terms of structured refactoring and anti-pattern detection. Furthermore, the LLM use cases spread from enterprise software and IoT systems to large-scale mobile and embedded systems, where Java is the most commonly used programming language in such studies. However, LLMs also suffer from issues such as accuracy and hallucinations, with other broader issues that need to be addressed in the future. 

Our study systematically synthesizes the current practice of LLM adoption in the software architecture domain, which shows clearly that LLM can contribute greatly to helping software architects in various aspects. It is optimistic that LLM, with fast-paced iterative updates, can continue to contribute to this domain with even more astonishing outcomes. 

\section*{Acknowledgment}
The research presented in this article has been partially funded by the Business Finland Project 6GSoft, by the Academy of Finland project MUFANO/349488 and by the  National Science Foundation (NSF) Grant No. 2409933.

\section*{Data Availability Statement}
We provide our raw data, and the MLR workflow in our replication package hosted on Zenodo\footnote{\url{https://doi.org/10.5281/zenodo.15032395}}.

\section*{Declaration of generative AI and AI-assisted technologies in the writing process}
During the preparation of this work, the author used ChatGPT to improve language and readability. After using this service, the authors reviewed and edited the content as needed and take full responsibility for the content of the publication.

\bibliographystyle{model1-num-names}
\bibliography{main}
% \bibliographystyleOS{model1-num-names}
% \bibliographyOS{OriginalStudies}
\bibliographystyleWL{model1-num-names}
\bibliographyWL{WhiteLitterature}
\bibliographystyleGL{model1-num-names}
\bibliographyGL{GrayLitterature}

\end{document}